%% file: ms.tex
\documentclass[12pt,preprint]{aastex}
\newcommand{\lam}{$\lambda$}
\newcommand{\Atm}{O$_2$ $b\,^1\Sigma^{+}_{g}$--$X\,^3\Sigma^{-}_{g}\,$}
\shorttitle{Nebular $s$-Process Abudances}
\shortauthors{Sharpee et al.}
\begin{document}

\title{$s$-Process Abundances in Planetary Nebulae}

\author{Brian Sharpee\altaffilmark{1}, Yong Zhang\altaffilmark{2,3},
Robert Williams\altaffilmark{2}, Eric Pellegrini\altaffilmark{4},
Kenneth Cavagnolo\altaffilmark{4}, Jack A. Baldwin\altaffilmark{4},
Mark Phillips\altaffilmark{5}, and Xiao-Wei Liu\altaffilmark{3}}

\altaffiltext{1}{SRI International, 333 Ravenswood Ave.  Menlo Park,
CA 94025, USA}
\altaffiltext{2}{Space Telescope Science Institute, 3700 San Martin
Drive, Baltimore, MD 21218, USA} 
\altaffiltext{3}{Department of
Astronomy, Peking University, Beijing 100871, P. R. China}
\altaffiltext{4}{Department of Physics \& Astronomy, Michigan State
University, East Lansing, MI 48824, USA} 
\altaffiltext{5}{Las Campanas Observatory, Carnegie Observatories,
Casilla 601, La Serena, Chile}

\begin{abstract} 

The $s$-process should occur in all but the lower mass progenitor stars
of planetary nebulae, and this should be reflected in the chemical
composition of the gas which is expelled to create the current
planetary nebula shell. Weak forbidden emission lines are expected
from several $s$-process elements in these shells, and have been
searched for and in some cases detected in previous investigations.
Here we extend these studies by combining very high signal-to-noise
echelle spectra of a sample of PNe with a critical analysis of the
identification of the emission lines of Z$>$30 ions.  Emission lines
of Br, Kr, Xe, Rb, Ba, and Pb are detected with a reasonable degree of
certainty in at least some of the objects studied here, and we also
tentatively identify lines from Te and I, each in one object.  The
strengths of these lines indicate enhancement of $s$-process elements
in the central star progenitors, and we determine the abundances of
Br, Kr, and Xe, elements for which atomic data relevant for abundance
determination have recently become available.  As representative
elements of the ``light'' and ``heavy'' $s$-process peaks Kr and Xe
exhibit similar enhancements over solar values, suggesting that PNe
progenitors experience substantial neutron exposure.

\end{abstract}

\keywords{ISM: abundances --- nuclear reactions, nucleosynthesis,
abundances --- planetary nebulae: general}

\section{Introduction}
%1
As remnants of stars that have evolved through the asymptotic giant
branch (AGB) phase, most planetary nebulae (PNe) are believed to
consist of material that has undergone nuclear processing in the
precursor star via the $s$-process.  The analysis of nebular emission
from elements that have experienced nucleosynthesis in the parent star
provides valuable information for stellar models.  However, the
detection of emission lines from ions enhanced by the $s$-process has
been hampered by their weakness and by uncertainties in the atomic
data needed for the analysis, including line wavelengths.

An initial attack on this problem was made a decade ago by
\citet{PB94} (hereafter PB94), who obtained a deep optical spectrum of
the high-ionization PN NGC~7027.  Using the best atomic data available
for the energy levels of the more prominent ionization stages of
elements in the fourth and fifth rows of the periodic table, they
identified a number of post-Fe peak emission lines in the nebula.
From the observed line intensities they concluded that the elements Kr
and Xe, the latter normally a predominantly $r$-process element in
stars of solar metalicity, were enhanced in NGC~7027 by factors of
$\sim$20 relative to their initial formation abundances, presumably
roughly solar.  They also detected \ion{Ba}{2} and [\ion{Br}{3}]
emission at intensities indicating that Ba could be enhanced whereas
Br might be depleted relative to solar values, subject to uncertainty
due to poorly known excitation cross sections.

Dinerstein and collaborators have subsequently pursued the study of
$s$-process abundances in PNe through surveys to detect IR
fine-structure nebular emission from post-Fe peak ions
\citep{D01,SD04} and from far-UV resonance-line absorption by
$s$-process elements in the intervening nebular shell seen in {\it
FUSE} spectra of their central stars \citep{SDB02,SD03}.  They
detected IR emission from Se and Kr in roughly 50$\%$ of their sample
PNe from which abundances of up to 10 times solar were derived for
these elements.  The {\it FUSE} spectra revealed \ion{Ge}{3}
absorption in five PNe for which Ge enhancements of 3-10 times solar
were deduced, clear evidence that central star AGB progenitors are
major sites of $s$-process element production.

As part of our on-going program to detect and identify the weakest
lines in PNe at visible wavelengths down to levels significantly below
that of the continuum, viz., $<10^{-5}$ the intensity of H$\beta$, we
have obtained very high signal-to-noise spectra at high spectral
resolution of some of the higher surface brightness nebulae.  Some
fraction of the weaker lines observed are likely to originate from
post-Fe peak elements and therefore we have used the automatic line
identification routine EMILI \citep{S03}, supplemented by recent
energy level data for heavier elements, to assist in the search for
such lines.  Since the publication of PB94, new calculations have been
made for spontaneous emission coefficients \citep{Bi95} and collision
strengths \citep{Sc97,SB98} for fourth and fifth row elemental ion
transitions.  In this study, we use these newer atomic parameters to
compute ionic and overall elemental abundances for Br, Kr, and Xe in
order to make comparisons with their abundances in \ion{H}{2} regions
and the sun, and to derive $s$-process enrichment factors relevant to
the study of $s$-process nucleosynthesis in the progenitor stars.

\section{Observations and Data Reduction}
%2
Our present sample of objects consists of four PNe (IC~2501, IC~4191,
NGC~2440, and NGC~7027) which satisfy the most important criteria for
the detection of weak emission lines in having (a) high surface
brightnesses, (b) low expansion velocities, and (c) except for
NGC~7027, relatively small internal dust extinction.  The latter two
criteria produce sharper lines and higher peak intensities relative to
the continuum.

We obtained spectra of the PNe IC~2501, IC~4191, and NGC~2440 during
two observing runs of two nights each in early 2003 using the Las
Campanas Observatory (LCO) Baade 6.5m telescope with the MIKE echelle
spectrograph.  Similar instrumentation setups were used during the two
runs.  MIKE is a dual-beam spectrograph which simultaneously measures
separate red and blue spectra, giving useful data over the continuous
wavelength ranges 3280--4700 \AA\ and 4590--7580 \AA\ at resolutions
($\lambda/\Delta\lambda$) of 28,000 and 22,000, respectively, for the
1 arcsec slit width we used.

A series of long and short exposures were taken of each nebula on one
or more nights, typically adding up to a period of 2-3 hours at a time
spent observing each object.  Because MIKE is used without an image
rotator at a Nasmyth focus, the orientation of the slit with respect
to each nebula rotated by a large amount during these series of
exposures.  The central star was placed along a line perpendicular to
the spectrograph slit so that the slit center was roughly midway
between the central star and outer edges of the nebula, and then the
telescope was guided to keep the star in that same position as seen on
the acquisition/guide TV. The result was that during the course of the
observations the slit swept out an arc about the central star, so that
the final spectrum integrates over an area of the nebula that is much
larger than the 1$\times$5 arcsec slit.  The total integration times
in the combined long exposures were 630 min for IC 2501, 450 min for
IC 4191, and 330 min for NGC 2440.  A journal of the observations is
given in Table~\ref{tab1}, and includes the approximate position of
the center of the slit with respect to each of the central stars at
the start of the series of exposures.

The optics of the MIKE spectrograph introduce strong distortion into
the image formed on the detector, so the projected emission lines have
significantly different tilts as a function of their position upon it.
When extracting spectra of objects such as these PNe which fill the
slit, the tilts introduce unacceptable smearing in the wavelength
direction (typically 3 pixels over the 40 pixel slit length) unless
they are corrected for during the extraction of the one-dimensional
spectra from the two-dimensional echelle image.  We wrote our own set
of auxiliary FORTRAN programs to calibrate the tilts and perform the
proper extraction. This procedure was tested on the emission lines in
the comparison lamp spectrum. Compared to the emission line from a
single pixel at the slit center, lines summed over the full slit
length came out at the same pixel location in the dispersion direction
and were broadened by 2 percent on average. We are thus confident that
the effects of these tilts are negligible in the extracted PNe
spectra. These tilt-corrected spectra were then fed into the same
suite of IRAF-based reduction programs used with the NGC~7027 data as
described next.

NGC 7027 was observed on 4 nights in June 2002 with the Mayall 4m
Telescope at Kitt Peak National Observatory using the Cassegrain
echelle spectrograph.  Because we were interested in detecting faint
40--50 km s$^{-1}$ FWHM emission lines over the widest possible
wavelength range, rather than in detailed measurements of the line
profiles, we used the short UV camera with the 79.1 grooves mm$^{-1}$
echelle grating, cross-dispersing grating 226-1 in first order, and a
GG-475 order separating filter.  This gave full wavelength coverage
from 4600 to 9200 \AA\ at $\sim$20 km s$^{-1}$ FWHM
($\lambda/\Delta\lambda=15,000$) resolution with our 2 arcsec slit
width, with partial coverage out to $\sim$9900 \AA.  On each night we
offset to the same position with the slit at PA 145$^\circ$ and
centered 0.5 arcsec N and 3.5 arcsec W of the central star, which is
the brightest part of the PN shell in H$\alpha$ emission.  Most of the
observing time was spent obtaining sequences of 1200s exposures which
added up to a total of 680 min of integration time.  We also took a
number of shorter exposures 30, 60, 120, and 300s in length to measure
bright emission lines that were saturated on the longer exposures.

The data for all four PNe were reduced using standard procedures with
IRAF-based reduction packages in the same manner that has been
described in detail in our discussion of comparable spectra of the PN
IC~418 that we obtained previously with the CTIO 4m echelle
spectrograph \citep{S04}.  A correction was made for the presence of
Rowland ghosts near strong lines \citep{B00}.  To correct for flexure
and temperature drift effects in the spectrographs, comparison lamp
spectra were taken at roughly one hour intervals, and the wavelength
calibration for each PN spectrum was made using the comparison lamp
taken nearest to it in time.  The wavelength fits indicate a
1-$\sigma$ uncertainty in the wavelength scale that varies over each
echelle order, and is of the order 4--6 km s$^{-1}$ for all of our
spectra.  The spectra were flux calibrated using observations of
several spectrophotometric standard stars from \citet{H94}.

The final extracted flux and wavelength calibrated spectra of all the
PNe were used for our line identification and analysis, and are
available in FITS format as an electronic supplement to the present
article.

\section{Emission Line Selection}
\begin{figure}
\includegraphics[width=40pc]{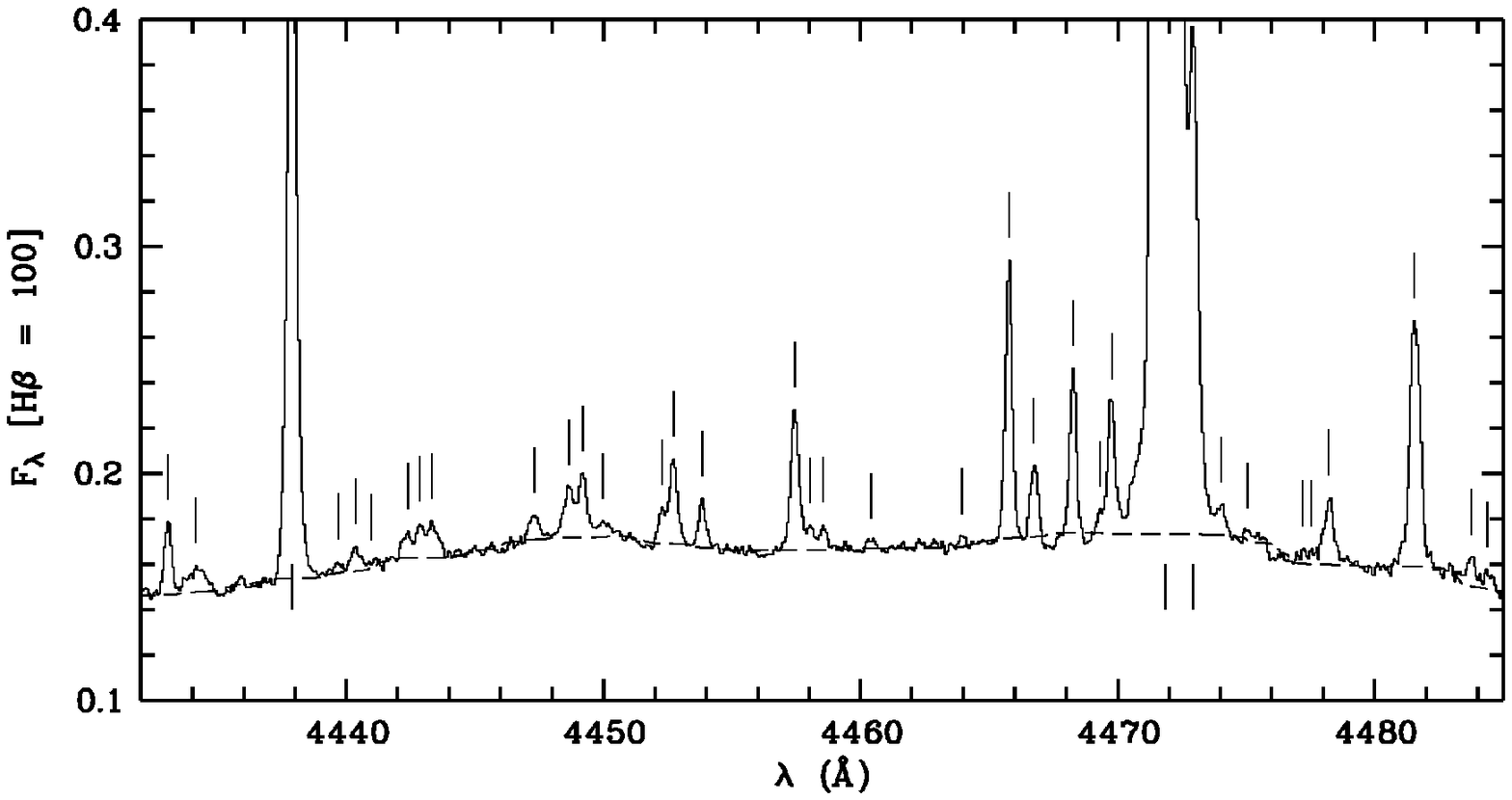}
\caption{Portion of the IC~2501 spectrum illustrating the automated
continuum fitting procedure.  The vertical lines designate emission
features defined by the ED fitting algorithm.}
\label{fig1}
\end{figure}

%3
The line selection procedure is initiated by fitting the continuum of
each echelle order of the final spectrum with a smooth function that
is pegged to the observed continuum at approximately ten wavelength
intervals distributed over each order.  We developed an automated
procedure that selects what should be reliable continuum points for
the fitting by selecting wavelength regions with a paucity of lines.
Because bad pixels, artifacts caused by scattered light in the
spectrograph due to strong emission lines, and line blends can cause
poor fits, the fits were reviewed manually for each order and adjusted
as necessary to insure that the continuum representation was valid.
An example of the output of the automatic continuum fitting algorithm
is shown in Figure~\ref{fig1}, where a typical fit prior to manual
adjustment is displayed.  Once the proper continuum level is
established, each order is sampled pixel by pixel to detect emission
features, which are defined as regions where the observed flux exceeds
the continuum flux by 7$\sigma$ or more over a wavelength interval
equal to or greater than that of the resolution of the spectrograph.

All putative emission features were examined individually by eye and
compared with their appearance on the original two-dimensional echelle
images to establish their reality.  We have found that all features
with fluxes greater than $12\sigma$ of that of the continuum, i.e.,
with S/N$>12$, are clearly visible on the echelle images.  Since
scattered light features usually trail across multiple orders one can
generally distinguish such artifacts from real lines by visually
examining the images.  The most uncertain aspect of the line detection
process is distinguishing multiple line blends in real emission
features.  The expansion velocities of PNe are such that intrinsic
line widths do vary by factors of 2-3 with the level of ionization,
so it can be difficult to discriminate between one broad line vs. two
or more closely spaced narrow lines.  Figure~\ref{fig1} gives an example
of the features within a selected wavelength region in one of our PNe
spectra that have been designated by our software as emission lines.

For detection of the weakest nebular lines, where the distinction
between a continuum noise spike and a real feature can be difficult to
establish with certainty, we have inter-compared the spectra of those
PNe that have similar levels of ionization but different radial
velocities.  Noise spikes do not generally produce features having the
width of the instrumental resolution, and instrument artifacts tend to
occur on the same place on the detector, which is at different rest
wavelengths in the PNe spectra because of their differing radial
velocities.  Since spectra were obtained on different instruments,
(MIKE and the 4-meter KPNO echelle), all significant scattered light
ghosts could be detected through comparisons between the PNe spectra
obtained with MIKE and the NGC~7027 spectrum.  Our final line lists do
contain a few weak features having S/N$<$7 which are present in at
least two of the objects.

The telluric nightglow emission spectrum was also sampled by these
spectra, as sky subtraction in these extended objects was deemed
impractical, particularly in regards to the subtraction adding
significant additional noise.  However, most nightglow lines were
distinguishable on the two-dimensional spectra by their uniform
intensity and characteristic shape and size, namely those of the
imaging slit.  The positions and intensities of prospective nightglow
lines were compared to those listed in the telluric feature atlases of
\citet{O96} and \citet{H03}, and likely matches removed from the
nebular list except in those cases where their blending with stronger
nebular lines was deemed only a minor contaminant.  For lines
considered as candidate $s$-process ion transitions, nightglow sources
were given additional scrutiny through a comparison with the
original spectra of the \citet{H03} atlas, the likely
identifications of the atlas lines \citep{C06}, time-averaged observed
intensities of their constituent systems \citep{CS06}, and in some
cases model spectra of those systems constructed with the molecular
simulation code DIATOM
(http://www-mpl.sri.com/software/DIATOM/DIATOM.html).
 
Following the selection of sets of likely nebular lines in each of the
PNe, their wavelengths and intensities were determined by single or
multiple, in the case of blended features, Gaussian fitting of their
profiles and immediate underlying continua.  Simple summing was used
for the most irregular and ill-defined line profiles with
intensity-weighted centroids utilized as wavelengths.  Emission lines
appearing in multiple orders were then collated and their attributes
averaged together.  Emission line intensities and wavelengths from
short and long duration exposure spectra of each PN were then
normalized to a particular fiducial frame.

Line wavelengths were shifted to the nebular rest frame through a
comparison with either the Balmer (MIKE PNe) or Paschen (NGC~7027)
sequence laboratory wavelengths.  Line intensities were de-reddened
with the Galactic extinction law of \citet{H83}, utilizing an
iterative process involving the magnitude of either the Balmer or
Paschen jump, and the Balmer or He II 5-n sequence decrements to
establish electron temperatures and c(H$\beta$) logarithmic extinction
at H$\beta$ values.  The c(H$\beta$) values for each PN are in good
agreement with those listed in \citet{CKS92}.  Errors in the final
line intensities in all PNe spectra, as determined from the formal
errors to the profile and continua fits, are similar to those found
for NGC~7027: 41\% for I$<$10$^{-5}$ I(H$\beta$), 20\% for
I=10$^{-5}$--10$^{-4}$ I(H$\beta$), 11\% for I=10$^{-4}$--10$^{-3}$,
and $>$6\% for I$>$10$^{-3}$ I(H$\beta$) on average.  This is
independent of any errors arising from the reddening correction.  For
weak lines, where the greatest contributor to uncertainty is the
indeterminate level of the true continuum, the formal errors in the
fit probably understate the actual uncertainty in the intensity.  From
random inspection of several lines at the 10$^{-5}$ H$\beta$ intensity
level, it is likely that some lines may have uncertainties ranging
upwards to 100\% in regions where the continuum level is rapidly
undulating, complicated by scattered light artifacts from adjacent
orders, or affected by strong telluric absorption.  Final geocentric
offsets and c(H$\beta$) values are presented in Table~\ref{tab2}

\begin{figure}
\plotone{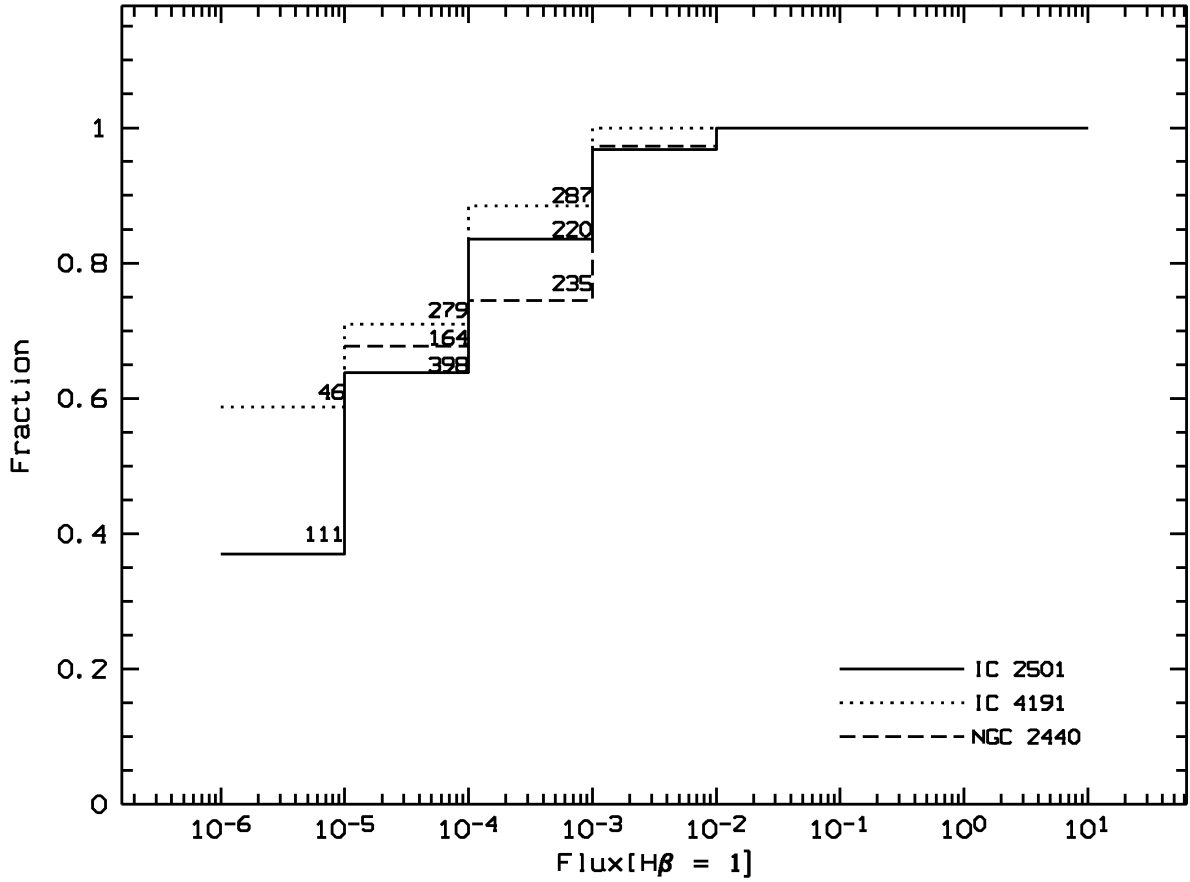}
\caption{The fraction of lines present in the spectrum of each PN
which are also detected in another of the PNe.  The numbers give the
total number of detected lines in each flux interval.}
\label{fig2}
\end{figure}

\begin{figure}
\plotone{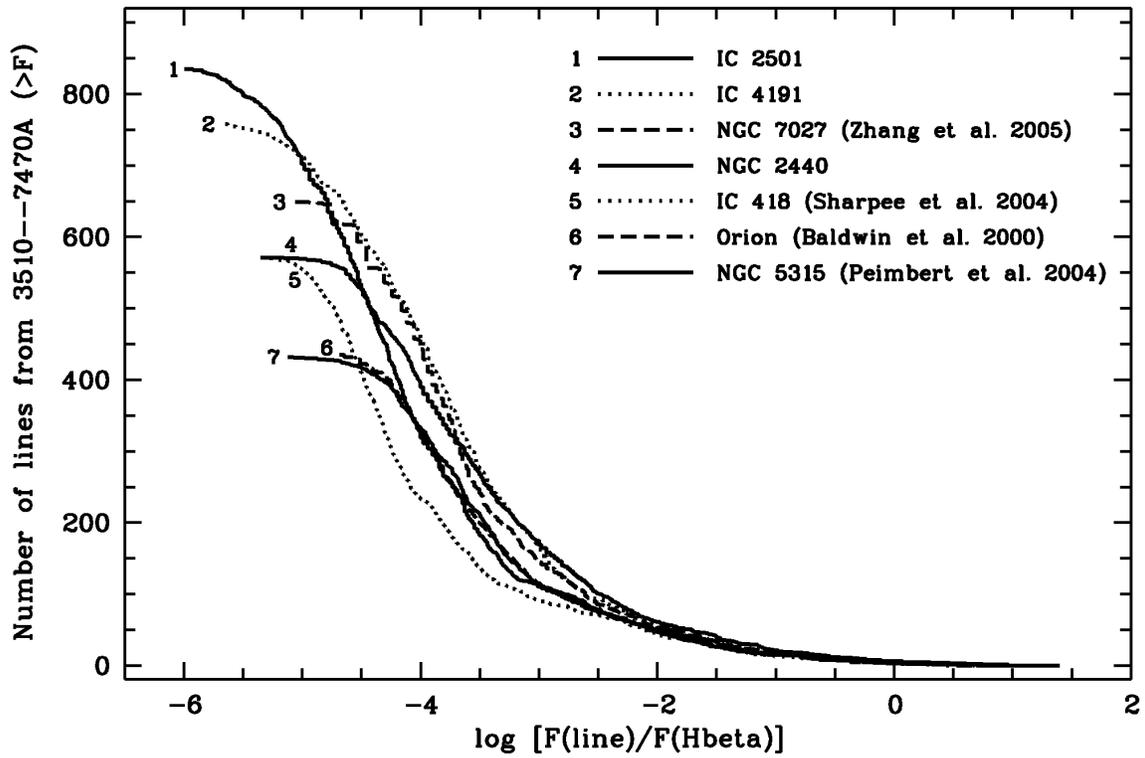}
\caption{The cumulative number of observed lines exceeding a given
flux level for recently published nebular spectra.  We consider only
those lines within the wavelength range 3510--7470~{\AA}, which is
covered by all the spectra.  }
\label{fig3}
\end{figure}

Figure~\ref{fig2} presents a histogram showing the fraction of lines
in each of the nebulae observed with MIKE for different intensity
levels relative to H$\beta$ that we have determined to be real and
which also appear in at least one of the other two PNe (the NGC~7027
spectrum is omitted here due to its different wavelength coverage and
spectral resolution).  The strongest emission lines are all detected
in each of the nebulae.  Even at intensities down to 10$^{-5}$
H$\beta$, where S/N$\sim$10 typically, roughly 50\% of the weak
features identified in the individual PNe also appear in one of the
other nebulae, suggesting that they are true nebular lines.  The
present spectra are among the deepest emission spectra taken of
nebulae, revealing some of the weakest lines yet observed.  This can
be seen in Figure~\ref{fig3}, where the cumulative number of lines
exceeding a given flux level relative to H$\beta$ is shown for several
of the most extensive PNe spectral studies published in the recent
literature.

\section{Plasma Diagnostics and Abundances}
%4
Electron temperatures and densities are presented in Table~\ref{tab3}.
The IRAF \textit{nebular} package task \textit{temden} \citep{SD95}
was used to equate relative intensities of collisionally excited
diagnostic lines to densities and temperatures by matching a
diagnostic for each from ions of similar ionization energy, and
solving for self-consistent values.  Each ion was modeled by a five or
greater level atom for this purpose, with spontaneous emission
coefficients and collision strengths for the five lowest levels drawn
primarily from the compilation of \citet{M83}.  Although these are not
the default values currently distributed with \textit{nebular}, this
atomic data set yielded good agreement in temperature and density
among diagnostics in our previous analysis of IC~418, and were
utilized in the original \textit{nebular} release.  Spontaneous
emission coefficients from \citet{FFT04} and collision strengths from
\citet{WB02} were used for \ion{N}{1} and \ion{Cl}{2} respectively.
For the remaining energy levels and for the ions \ion{Cl}{4} and
\ion{K}{5}, the atomic data used in the most recent release of
\textit{nebular} was utilized.  The departure of the [\ion{K}{5}]
density diagnostic from other density diagnostic values, particularly
in NGC~2440 and IC~4191, may indicate errors in this atomic data.
However, $s$-process elemental abundance derived later using this
diagnostic (Kr$^{+4}$/H$^+$ and Xe$^{+5}$/H$^+$) were completely
insensitive to the utilized density.  Errors in diagnostic values were
determined by selecting the extrema values from computations at all
combinations of diagnostic line ratios plus and minus their
(1-$\sigma$) uncertainties, including an error estimate for the
reddening correction.  Indeterminate error limits, such as occurred
when a ratio value exceeded the asymptotic limit or where the paired
diagnostics failed to converge at a particular ratio value, are listed
without a value. Balmer (MIKE PNe) and Paschen (NGC~7027) jump
temperatures were calculated in the manner described by \citet{Z04},
and as reported by them for many PNe, they are lower than those
temperatures determined from the collisionally excited lines.

Ionic abundances derived from both collisionally excited and nominal
radiative recombination lines are presented in Table~\ref{tab4}.  The
IRAF \textit{nebular} package tasks \textit{ionic} and
\textit{abundance} were used to make computations from the
collisionally excited lines listed in Table~\ref{tab3} using
temperatures and densities derived from the diagnostics with the
closest ionization potential to each ion.  Where no diagnostics where
clearly appropriate, averaged values for temperature and density where
used.  Uncertainties were computed in the same manner as the
diagnostic values, by calculation of the abundance at every
combination of temperature, density, and line intensity ratio plus or
minus their (1-$\sigma$) uncertainties (where available) and selecting
the extrema values.  Uncertainties were not calculated for abundance
computed with averaged diagnostics values.  For the recombination
lines, effective recombination coefficients were combined with line
intensities to make abundance determinations, in the manner described
by \citet{S04}, with temperature and densities again drawn from
diagnostics nearest in ionization potential to each ion.  For
He$^+$/H$^+$, the coefficients of \citet{S96} or \citet{BSS99} were
used to determine an average abundance from the \lam4923, \lam5876,
and \lam6678 lines, with corrections for collisional excitation (Case
A triplets, Case B for singlets) taken from \citet{KF95}, while
coefficients from \citet{SH95} were used to calculate an average
He$^{+2}$/H$^+$ abundances from various \ion{He}{2} lines.  To
determine C$^{+2}$, N$^{+2}$, O$^{+2}$, and Ne$^{+2}$ abundances
relative to H$^+$, recent effective recombination coefficients
\citep{S94,K98,DSK00,KS02} for the strongest observable multiplets
were used to calculate abundances.  As seen in Table~\ref{tab4},
O$^{+2}$/H$^+$ and Ne$^{+2}$/H$^+$ abundances deduced from the
collisionally excited lines are systematically lower than those
derived from radiative recombination lines, as is generally the case
in PNe \citep{L04,RG05}.

\section{Emission Line Identification}
%5
The emission line identification code EMILI \citep{S03} was used to
make the majority of emission line identifications.  EMILI creates
models of the ionization-energy dependent velocity field and
ionization level of a PN or H II region from user-supplied empirical
data.  These models are used by EMILI to select from a large atomic
transition database all transitions within five times an observed
line's wavelength measurement error, and to compute relative
intensities for emission lines corresponding to those transitions.
Those transitions predicted to produce emission line intensities
within 3 orders of magnitude of the highest value among all
transitions initially selected are then subjected to a test for the
presence of lines from the same LS-coupled multiplets (but not all
lines from the same upper level) at expected wavelengths and relative
intensities.  Potential identifications are then assigned a numeric
likelihood parameter based on their wavelength agreement with the
observed line, strength of the predicted emission line, and results of
the multiplet check, and are ranked and presented.
  
Many emission lines from Z$>$30 elemental ions were observed in
NGC~7027 by PB94.  To place these lines on an equal footing for
identification purposes with those arising from more abundant lighter
elements, EMILI used the Atomic Line List v2.05 of P. van
Hoof\footnote{http://www.pa.uky.edu/\~peter/newpage/} as its reference
database, which extends to Z=36 (Kr).  The latest experimental
determinations available in the literature for ground electron
configuration energy levels of Z$>$36 ions were then added to this
database.  Transition wavelengths for the mostly optically forbidden
transitions among these levels were constructed by differencing the
level energies.  Sources for all Z$>$30 ion energy levels and
associated atomic data used in subsequent analysis are provided in
Table~\ref{tab5}.  All ions with certain or probable line
identifications in NGC~7027, as rated by PB94, most of those with more
tenuous identifications, and a handful of other ions they suggested to
be worthy of future consideration at a higher spectral resolution,
were incorporated into the database.  However, ions with level
uncertainties perceived or explicitly stated to be greater than 1.0
cm$^{-1}$ in their source literature were excluded.

EMILI was run against each nebula's set of observed emission line
wavelengths and intensities.  Electron temperature and density values
were provided to EMILI by the results of standard plasma diagnostics,
derived from strong collisionally-excited lines with certain
identifications.  Construction of the empirical kinetic and ionization
models were also drawn from those same lines.  The EMILI reference
elemental abundances were set to the solar values of \citet{L03}.
EMILI was also run against the IC~418 line list of \citet{S03}, to
determine if some of its remaining unidentified lines could be
identified with an expanded set of transitions, and to also act as a
foil for the mostly higher excitation PNe considered in the present
sample.  The line identification lists produced by EMILI were then
visually inspected, and the EMILI-preferred assignments were compared
to identifications in the literature for those same lines in PNe and H
II region spectra of similar depth, spectral resolution, and level of
ionization.  The entries in our final line list are in many cases
taken from among the highest-ranked EMILI suggested identifications.

A segment of a final line list is given in Table~\ref{tab6}; the full
line list is available in the electronic version of this manuscript.
This table lists all observed lines with their characteristics, viz.,
observed wavelength in the nebula rest frame, identifications, full
widths at half maxima, S/N, observed and reddening corrected
intensities relative to H$\beta$, and their most likely
identifications.  Lines with uncertain identifications are denoted by
a ``:'', likely blends with a ``bl'' or ``ns'' if blended with a
telluric emission feature, and uncertain features are marked with a
``?''.  Only the perceived strongest component of a blend is listed in
the table.  Lines without obvious identifications are listed here
without an identification.

\section{Z$>30$ Line Identifications}
\label{Z30id}
%6
PB94 detected 25 emission lines from several Z$>$30 ions in their
optical spectra of NGC~7027, including Se, Br, Kr, Rb, Sr, and Y from
the fourth row of the periodic table, Xe and Ba from the fifth row,
and Pb from the sixth.  Eighteen of these detections were considered
``certain'' or ``probable'', and 7 considered ``possible''.  Lines from 13
additional Z$>$30 ions were also either tentatively identified or
proposed as future targets for spectra with greater spectral
resolution and better S/N.

Given the depth and high-resolution of the spectra considered here,
confirmation of the PB94 identifications in multiple PNe was sought,
as were additional lines belonging to other Z$>$30 ions.  The use of
EMILI allows prospective Z$>$30 transitions and weaker transitions of
more abundant lighter elements to be treated equally for purposes of
emission line identification.  Collision strength and spontaneous
emission coefficient calculations for these transitions allow accurate
predictions to be made of the relative intensities of lines arising
from the same ion, allowing identifications of their lines to be made
with more confidence.

The line lists for the present PNe sample and IC~418 \citep{S03} were
searched for the strongest expected Z$>$30 ion transitions within
their observed bandpasses.  These primarily were the
$^3$P$_{1,2}$--$^1$D$_2$ nebular transitions of ions with 4p$^{2,4}$
and 5p$^{2,4}$ valance electrons, the
$^4$S$^{o}_{3/2}$--$^2$D$^{o}_{3/2,5/2}$ nebular transitions of ions
with 4p$^3$ and 5p$^3$ valence electrons, and the
$^2$P$^{o}_{1/2,3/2}$--$^2$P$^{o}_{1/2,3/2}$ fine-structure
transitions for 4p, 4p$^5$, 5p, 5p$^5$, and 6p valance electron ions.
Fine structure $^3$P--$^3$P transitions, when accessible in the
visible, were also included.  For the cases of \ion{Kr}{4},
\ion{Xe}{4}, and \ion{Br}{4}, auroral
$^2$D$^{o}_{3/2,5/2}$--$^2$P$^{o}_{1/2,3/2}$ and trans-auroral
$^4$S$^{o}_{3/2}$--$^2$P$^{o}_{1/2,3/2}$ transitions were also
considered.  The permitted resonance lines of \ion{Ba}{2} and
\ion{Sr}{2} were also searched for.  Initially, all wavelength
coincidences of 1\AA\ or less between an observed line and a
transition wavelength were considered possible Z$>$30 lines,
regardless of the identifications recommended by EMILI for those
lines.

Tables~\ref{tab7}--~\ref{tab10} present excerpts from the EMILI output
for observed lines believed to represent the best cases for a Z$>$30
transition as the actual identification for an observed line in at
least one of the PNe in which that line appeared.  For each observed
line in each PN, the identifications of highest rank, as specified by
the ``Identification Index'' or IDI, the EMILI figure-of-merit for
quality of an identification (5 or less is considered a quality
identification), and their multiplet search statistics (numbers
expected/observed, hereafter “multiplet statistics”) are included.
The IDI value is followed by a letter A$\rightarrow$D if among the top
four highest ranked transitions, with ``A'' being the highest.
Additional identifications drawn from the literature and from the
terrestrial nightglow (prefaced with ``SKY'') are also included.
Identifications that did not yield predicted line intensities within
three orders of magnitude of the strongest value among all
identifications, or were outside the 5-$\sigma$ search radius, do not
have a calculated IDI value.  The IDI value/rank is sometimes followed
by a symbol, an asterisk indicating the most likely single
identification, ``bl'' indicating components of a likely blend, and
``:'' indicating indeterminate alternate identifications.  The
reddening corrected intensity of the feature attributable to the
putative s-process identification, if corrected for any blending as
described in succeeding sections (due to other identifications denoted
with a ``bl''), is presented in italics below its originally observed
value.  The limiting intensity, determined from the local minimum flux
of line detection considered certain (S/N=7), or from imposition of
artifical lines of S/N=7 at that wavelength, is presented in brackets
for the case of s-process transitions without corresponding observed
features. An ``OUT'' label indicates features residing outside the
observed bandpass of a spectrum.  Finally, features of dubious reality
are denoted by a ``?'' following the observed wavelength.
 
\begin{figure}
\plotone{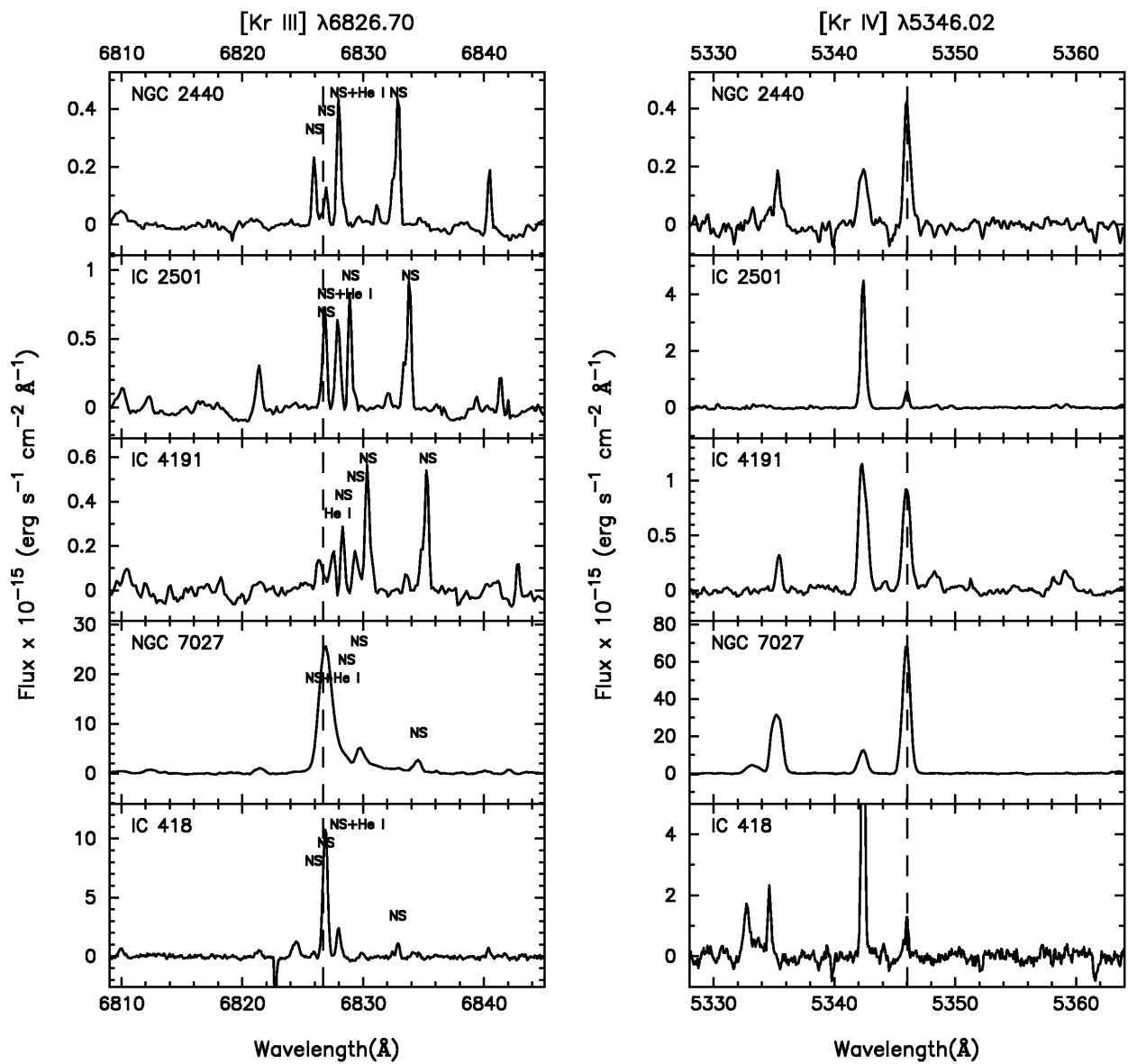}
\caption{Continuum subtracted spectra in the immediate vicinity of
several prominent Kr, Xe, and Br lines, shifted to each nebula's rest
frame.  Predicted wavelengths are indicated by dashed lines.  Labels
``NS'' indicate telluric nightglow lines mentioned in the text.  The
regions around [\ion{Kr}{3}] \lam6826.70, [\ion{Kr}{5}] \lam6256.06, and
[\ion{Xe}{3}] \lam5846.77 are expanded in subsequent figures and
discussed in the text.  \label{fig4}}
\end{figure}

Figure~\ref{fig4} depicts continuum-subtracted spectra, with
wavelengths shifted to the nebular rest frame, in the vicinity of
those Kr, Xe, and Br lines for which a positive identification was
made in at least one PN, and which were potentially observable in all
four PNe of the present sample and in IC~418.  Comments on
identifications pertaining to individual Z$>$30 elemental ion lines
follow.

\subsection{Kr Line Identifications}
%6.1
PB94 noted that [\ion{Kr}{3}] 4p$^4$ $^3$P$_2$--4p$^4$ $^1$D$_2$
\lam6826.70 has long been observed in various novae and PNe, but has
seldom been identified as such.  They identified [\ion{Kr}{3}] as a
strong line in NGC~7027, blended with weak \ion{C}{1} (V21)
\lam6828.12 and \ion{He}{1} 3s $^3$S--16p $^3$P$^o$ \lam6827.88 lines.

\begin{figure}[t]
\begin{center}
\plotone{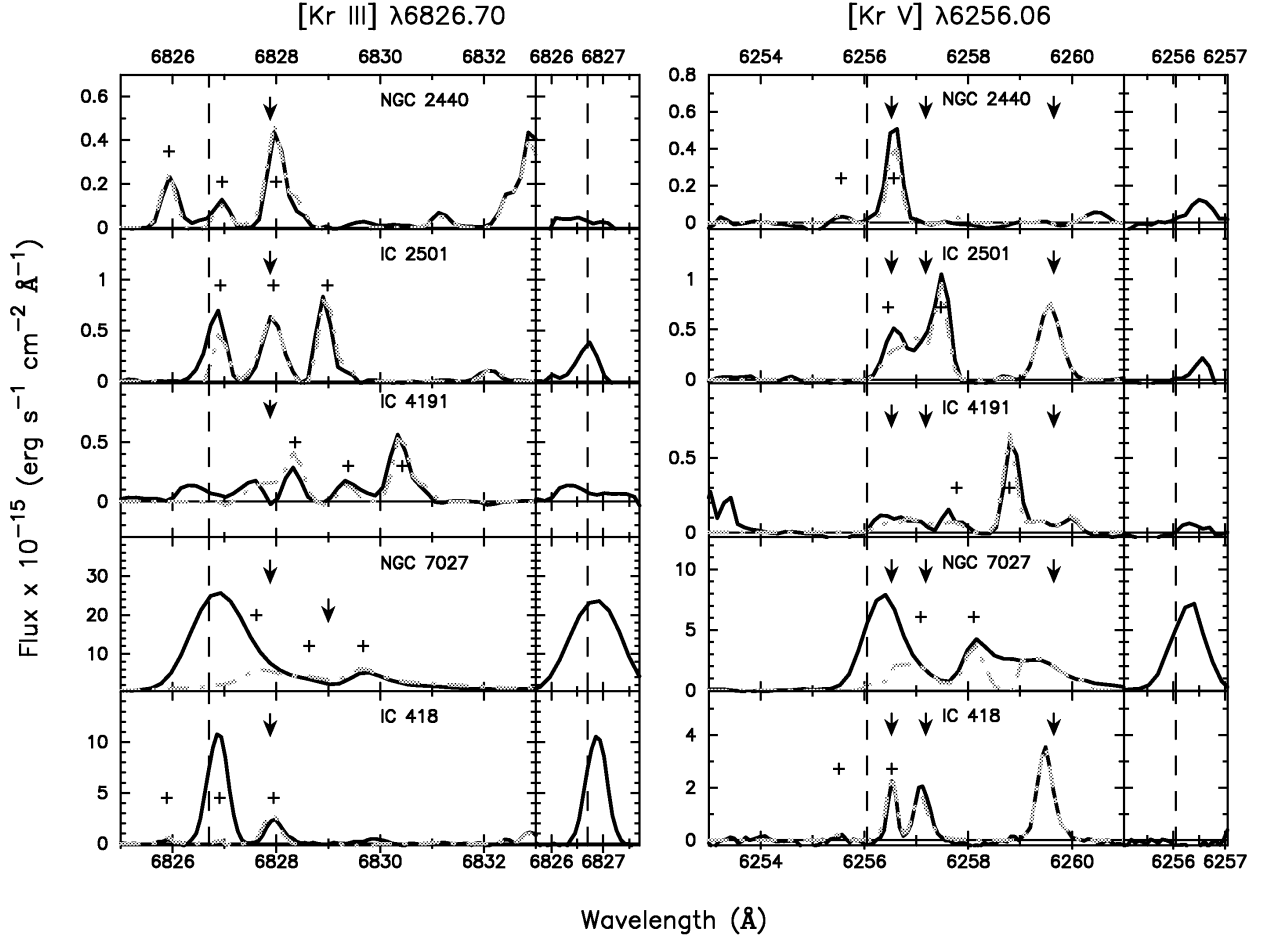}
\caption{LEFT: Spectra in the vicinity of [\ion{Kr}{3}] \lam6826.70
(black line) shifted to nebular rest wavelengths, with similarly
shifted and superimposed coadded modeled spectra of telluric emission
and observed interloping nebular features (dot-dashed gray line).
Pluses indicate telluric OH Meinel 7-2 R$_1$(3.5) \lam6827.46,
R$_1$(4.5) \lam6828.47, and R$_1$(2.5) \lam6829.49 lines (left to
right), nebular \ion{He}{1} \lam6827.88 is marked with an arrow, and a
\ion{O}{6} \lam1032 Raman feature at \lam6829.16 (NGC~7027) marked
with a lower arrow.  RIGHT: Same for [\ion{Kr}{5}] \lam6256.06,
with OH Meinel 9-3 Q$_2$(0.5) \lam6256.94 and Q$_1$(1.5) \lam6257.96
marked as pluses (left to right), and nebular \ion{C}{2} \lam6256.52,
\lam6257.18, \lam6259.65 lines marked with arrows (left to right).
Panels to right show residuals after subtraction in vicinity of
predicted wavelengths of target line (vertical dashed line in all
panels). \label{fig5}}
\end{center}
\end{figure}

As seen in Table~\ref{tab7} and Figure~\ref{fig5}, the higher
resolution of our present PNe spectra cleanly separates both the
\ion{He}{1} and \ion{C}{1} line from the putative [\ion{Kr}{3}] line,
except for NGC~7027, where \ion{He}{1} contributes minimally to its
red shoulder.  However, Figure~\ref{fig5} shows that the R branch head
of the telluric nightglow OH Meinel 7-3 band is a serious contaminant
in this region.  In NGC~7027, a broad \ion{O}{6} \lam1032 Raman
scattering line at \lam6829.16, seen previously in NGC~7027 by
\citet{Z05}, is also observed.  The OH contribution was represented by
a model of the band normalized in intensity to the uncontaminated
R$_1$(1.5) line at \lam6834.01.  The \ion{He}{1} 3s $^3$S--15p
$^3$P$^o$ \lam6855.91 line profile was shifted and scaled by a factor
of 0.83 to represent \ion{He}{1} \lam6827.88 (Smits et al.\ 1991; Case
B, for the most appropriate grid point: T$_e$=10$^4$K, n$_e$=10$^4$
cm$^{-3}$).  The profile of the companion \ion{O}{6} Raman scattering
line observed at \lam7088, scaled upwards by a factor of four to yield
the best fit with the red wing of the putative [\ion{Kr}{4}] feature,
represented the \lam6829.16 line.  These line profiles and the
telluric model were subtracted from the continuum normalized spectra
to yield the residual features shown in Figure~\ref{fig5}.

For all PNe except NGC~2440, where the observed line appears to be
entirely comprised of the OH line, there appears to be a substantial
residual feature that is coincident, but except for IC~4191, slightly
to the red of the predicted wavelength for [\ion{Kr}{3}] \lam6826.70.
The EMILI results for the corresponding observed lines suggest that
besides the \ion{C}{1} and \ion{He}{1} transitions mentioned earlier,
the [\ion{Fe}{4}] \lam6826.50 feature is probably the only other
sensible alternative identification.  However, given the absence of
other multiplet members that should be observed with this line and the
relative observed strengths of the likely strongest [\ion{Fe}{4}]
lines \citep{R03}, this is not a likely identification except perhaps
for IC~4191.  Unfortunately, the companion [\ion{Kr}{3}]
$^3$P$_1$--$^1$D$_2$ line at \lam9902.3, predicted to be $\sim$18
times weaker by \citet{BH86b}, is only potentially observable in
NGC~7027, where a line listed at \lam9903.55 is much too strong and
most likely associated with \ion{C}{2} 4f $^2$F$^o$--5g $^2$G
\lam9903.67.  Nevertheless, we conclude that the [\ion{Kr}{3}]
\lam6826.70 identification is relatively certain in IC 418, IC 2501
and NGC 7027, uncertain in IC 4191, and that line is not observed in
NGC 2440.

Examination of Table~\ref{tab7} shows that both transitions of the
[\ion{Kr}{4}] $^4$S$^o$-$^2$D$^o$ nebular multiplet \lam\lam
5346.02,5867.74 are clearly identified in every spectrum, appearing as
the primary EMILI ranked identifications.  For \lam5346.02, the
[\ion{Fe}{2}] \lam5347.65 identification is clearly too far away and
the \ion{S}{2} (V38) \lam5345.71 dielectronic transition is unlikely
given the absence of other multiplet members.  The \ion{Al}{2}
\lam5867.8 identification for \lam5867.74, given by \citet{B00} and
\citet{S03}, is clearly superseded.  With a theoretical intensity
ratio I(\lam5346.02)/I(5867.74) of 0.65 \citep{Sc97}, assuming
electron densities well below the critical density values of
$\sim$18--1.3$\times 10^7$ cm$^{-3}$ for the $^2$D$^{o}_{3/2}$ and
$^2$D$^{o}_{5/2}$ levels respectively, as justified from the values in
Table~\ref{tab5}, the observed \lam5346.02 intensity appears to be
somewhat too high for IC~4191 and IC~418.  This suggests the possible
blending of [\ion{Kr}{3}] \lam5346.02 with another transition, perhaps
\ion{C}{3} (V13.01) \lam5345.85, although poor multiplet statistics
and the weakness of the strongest \ion{C}{3} lines in IC~418 cast
doubt on this possibility for that PN.  However, the other three PNe
have ratios of 0.76 (NGC~2440), 0.72 (IC~2501), and 0.73 (NGC~7027),
which are slightly higher but consistent with the expected value
within the combined measurement errors of the line intensities.

The auroral [\ion{Kr}{4}] $^2$D$^{o}_{3/2,5/2}$-$^2$P$^{o}_{3/2}$
\lam\lam6107.8,6798.4 identifications are less straightforward,
except for NGC~7027 where both are considered certain according to
both PB94 and the present EMILI results.  In NGC~2440, the observed
line corresponding to the \lam6798.4 transition, which should be 2.5
times weaker than \lam6107.8 \citep{BH86b}, is actually stronger,
while in IC~2501 the stronger \lam6107.8 line is not present at all.
As suggested by both PB94 and the EMILI results, \ion{C}{2} (V14)
\lam6798.10 may be responsible for either the total intensity in
IC~2501 or excess intensity in NGC~2440.  While PB94 note that
\ion{C}{2} (V14) is of dielectronic recombination origin and its
multiplet components should not have relative intensities expected due
to LS-coupling rules, an examination of the intensities of all lines
from this multiplet appearing in each PN suggests that the intensities
do appear to roughly follow those rules with the exception of what
would be the strongest line at \lam6783.91.  Therefore, as an
approximation, a LS-coupled relative strength of \lam6798.10 to
\lam6791.47 of 0.16, instead of the one-to-one ratio with \lam6812.28
used by PB94, was utilized to correct the putative [\ion{Kr}{4}]
\lam6798.4 feature in both NGC~2440 and NGC~7027, as \lam6812.28
appears itself appears in only one of the PNe spectra (it would be the
weakest feature in the multiplet if LS-coupling rules held).
Subsequent abundance analysis also suggests that the [\ion{Kr}{4}]
\lam6107.8 line itself may be too strong relative to its nebular
counterparts in NGC~2440 and IC~4191, admitting [\ion{Fe}{2}]
\lam6107.28 as an alternate identification that appears more likely in
the latter PN.  In summary, both auroral transition can only be
identified comfortably in NGC~7027.

The third magnetic dipole transition of the [\ion{Kr}{4}] auroral
multiplet $^2$D$^{o}_{3/2}$--$^2$P$^{o}_{1/2}$ \lam7131.3 is not
definitively detected in any spectrum including NGC~7027.  However, in
the NGC~7027 spectrum, a feature is observed at \lam7131.65 with
intensity $7.7\times 10^{-5}$ I(H$\beta$), comparable to \lam6107.8.
It originally was assumed to be an under-corrected Rowland ghost
arising from the nearby saturated [\ion{Ar}{3}] $^3$P$_2$--$^1$D$_2$
\lam7135.80 line, but might instead be associated with \lam7131.3.
The electric quadrupole line [\ion{Kr}{4}]
$^2$D$^{o}_{5/2}$--$^2$P$^{o}_{1/2}$ \lam8091.0 is forty times weaker
and therefore undetectable.

PB94 claim a detection of [\ion{Kr}{5}] $^3$P$_1$--$^1$D$_2$
\lam6256.06 as a blend with \ion{C}{2} (V10.03) \lam6257.18 at their
instrumental resolution.  In our spectra, as is shown in
Figure~\ref{fig5}, \ion{C}{2} \lam6257.18 is resolvable from the
putative [\ion{Kr}{5}] feature in all spectra except NGC~7027, where
it slightly contributes to its red wing.  However, in IC~2501 and
NGC~7027, as also noted by \citet{Z05}, another \ion{C}{2} line,
dielectronic \ion{C}{2} (V38.03) \lam6256.52, is believed to be a
significant contaminant in some PNe based upon the strength of the
nearby \lam6250.76 line from the same multiplet and upon favorable
EMILI assessments.  The profiles of both \ion{C}{2} lines were
approximated and subtracted from all the PNe spectra in
Figure~\ref{fig5}, where the \ion{C}{2} (V10.03) \lam6259.56 line
profile scaled downward by a factor of 0.56 represents \lam6257.18,
and the \ion{C}{2} \lam6250.76 profile, also scaled downward by a
factor of 0.56 according to LS-coupling statistics, represents
\lam6256.52.  The contribution of the interloping telluric OH 9-3 band
was accounted for through subtraction of a model of the band
normalized in intensity to the nearby Q$_1$(2.5) \lam6265.21 line.

The resulting residual plots show a distinct line just to the red of
the predicted wavelength of [\ion{Kr}{5}] \lam6256.06 in NGC~2440,
IC~2501, and NGC~7027, that we believe can be definitely identified as
such in all three PNe.  For IC~4191, the residual flux, $3.0\times
10^{-6}$ I(H$\beta$), is probably too low to be an actual line, and
the entirety of the original profile is ascribed here to \ion{C}{2}
(V38.03) \lam6256.52, particularly since both the original line and
\lam6250.76 share the same broad profile in the spectrum.  In IC~418,
the putative [\ion{Kr}{5}] profile is completely removed by the
combination of the nightglow and \ion{C}{2} (V38.03) \lam6256.52 model
profiles, as is appropriate given the nebula's low excitation.

The other nebular line, [\ion{Kr}{5}] $^3$P$_2$--$^1$D$_2$
\lam8243.39, accessible only in the bandpasses of the NGC~7027 and
IC~418 spectra, sits amidst the head of the Paschen series, rendering
it difficult to disentangle at the resolution of our NGC~7027
spectrum.  In the IC~418 spectrum, a comparable observed line is
indistinguishable from other nearby lines in the Paschen series and is
clearly identifiable as \ion{H}{1} 3--43 \lam8243.39, with the
\ion{Kr}{5} identification unlikely due to the PN's low ionization
level.  Returning to NGC~7027, PB94 identified this transition at
\lam8242.7 as a blend with \ion{N}{1} (V2) \lam8242.39 and \ion{O}{3}
5g G $^2$[9/2]$^o$--6h H $^2$[11/2] \lam8244.10, compromised by
telluric absorption.  In our NGC~7027 spectrum, [\ion{Kr}{5}]
\lam8243.39 is tentatively identified as the blue peak at \lam8244.33
of a resolved two-line blend with \ion{H}{1} 3--42 \lam8245.64, and
which appears significantly affected by telluric absorption.  The
alternate identification for the line, \ion{N}{1} \lam8242.39, is
clearly resolved here as a separate feature.  Because this observed
line has a measured intensity ratio with \lam6256.06 roughly
comparable to what would be expected if both lines were due to
[\ion{Kr}{5}], I(\lam6256.06)/I(\lam8243.39)=1.1 \citep{BH86a}, it is
believed that \ion{O}{3} \lam8244.10 contributes only a minor amount
to the line, although the presence of the telluric absorption
complicates matters.  To summarize, both [\ion{Kr}{5}] lines appear to
be present with about the expected intensity ratio in the spectrum of
NGC~7027, but both are absent in the spectrum of IC~418. For the cases
of IC 2501 and NGC 2440, [\ion{Kr}{5}] \lam6256.06 is probably
present.

The spectra were also examined for evidence of other auroral and
trans-auroral transitions of Kr ions.  Only the auroral line
[\ion{Kr}{5}] $^1$D$_2$--$^1$S$_0$ \lam5131.78 satisfied the initial
$\pm$1\AA\ screening criterion and also appeared in the EMILI output
as a possible identification for an observed line at $\sim$5131.0\AA\
in all PNe.  The case for this identification is lessened by the
expected weakness of this line compared to other [\ion{Kr}{5}] lines
that were not definitively identified, and particularly because of its
high observed intensity of $1.3\times 10^{-3}$ I(H$\beta$) relative to
the $\sim 1\times 10^{-4}$ I(H$\beta$) for \lam\lam6256.06,8243.39 in
NGC~7027, the nebula that might be expected to exhibit the best
evidence for this line given the strength of the other confirmed Kr
lines.  Instead either \ion{O}{1} 3p $^3$P--8d $^3$D$^o$ \lam5131.25,
which was the highest ranked line in all but one nebula (IC~418) and
that has a strength comparable to other members of the same sequence,
or \ion{C}{3} 5g $^3$G--7h $^3$H$^o$ \lam5130.83 appear the more
likely identifications.

\subsection{Xe Line Identifications}
%6.2
A number of Xe ion transition identifications are also considered
probable in our spectra.  The EMILI statistics for Xe identifications
are affected by the low solar Xe abundance, which is an order of
magnitude lower than Kr \citep{L03}.  This contributes to low
predicted relative emission line intensities and consequently higher
or absent IDI values for its identifications as compared to those
computed for weak transitions from more abundant elements.  As such,
every appearance of a Xe ion transition as a candidate identification
for an observed line in an EMILI list, regardless of its IDI value,
was given serious consideration as it signaled that alternative
identifications from more abundant elements did not predominate
despite the advantage arising from greater abundances.
Table~\ref{tab8} lists the EMILI statistics for lines judged most
likely to correspond to Xe ion transitions, while Figure~\ref{fig4}
depicts the regions around the most likely observed transitions.

\begin{figure}[t]
\begin{center}
\includegraphics[width=20pc]{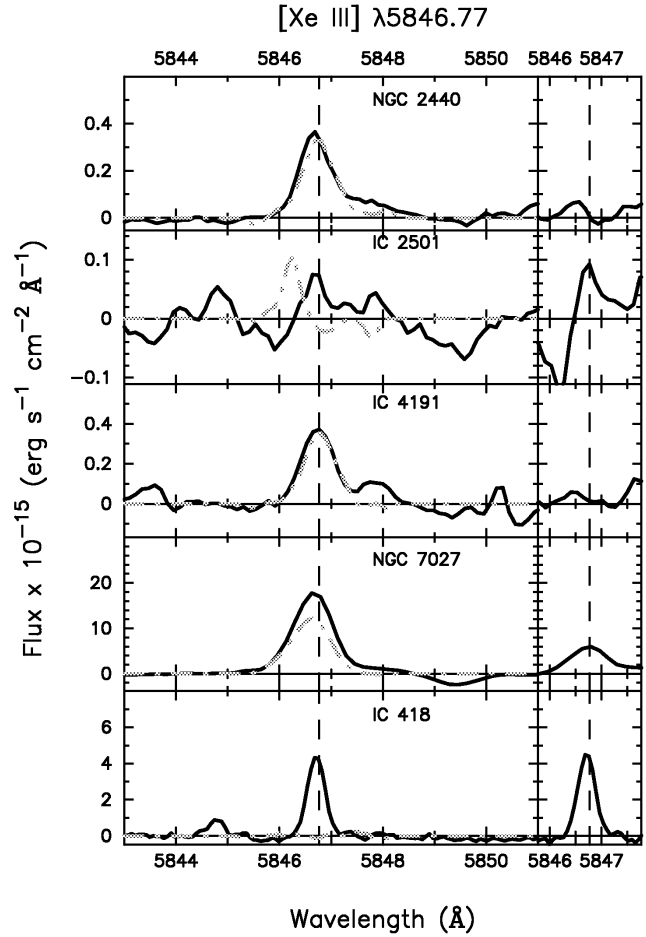}
\caption{Spectra in vicinity of [\ion{Xe}{3}] \lam5846.77 (black line)
shifted to nebular rest wavelengths, with superimposed
profile of the \ion{He}{2} 5-32 \lam5837.06 shifted to the rest
wavelength of the \ion{He}{2} 5-31 line and scaled as described in the
text (gray dot-dashed line).  Panels to right show residual spectra
after subtraction near the rest
wavelength of the [\ion{Xe}{3}] \lam5846.77 line (vertical dashed line
in all panels). \label{fig6}}
\end{center}
\end{figure}

The identification of [\ion{Xe}{3}] $^3$P$_2$--$^1$D$_2$ \lam5846.77
in NGC~7027, was given by PB94 to an excess intensity in the
\ion{He}{2} 5--31 \lam5846.66 line.  This excess was searched for in
the present PNe sample through the subtraction of the \ion{He}{2}
\lam5837.06 profiles, shifted to the \lam5846.66 line position and
scaled assuming I(\lam5837.06)/I(\lam5846.66)=0.92 (Storey and Hummer
1995; Case B, for the most appropriate grid point: T$_e$=10$^4$K,
n$_e$=10$^4$ cm$^{-3}$) as is shown in Figure~\ref{fig6}.  The
intensity of \ion{He}{2} \lam5837.06 was not corrected for the
presence of \ion{C}{3} 7h $^3$H$^o$--18i $^3$I \lam5836.70 because the
line used for this correction by PB94, \ion{C}{3} 7i $^3$I--18k
$^3$K$^o$ \lam5841.2, was either not present or was an identification
of low rank (IDI=6, ranked 5th) for a corresponding line in NGC~7027.
As seen in Figure~\ref{fig6}, distinct residual profiles are present
after subtraction of \ion{He}{2} \lam5846.66 for both NGC~7027 and
IC~418 (the latter having negligible \ion{He}{2}), and at the correct
wavelength for [\ion{Xe}{3}] \lam5846.77.  A bizarrely shaped profile
is seen for IC~2501, and no profiles are seen for either NGC~2440 or
IC~4191 after subtraction.  While the corresponding observed lines are
probably \ion{He}{2} \lam5846.66 in the latter two PNe, IC~2501 has
very weak \ion{He}{2} lines, suggesting that the residual profile is
attributable to something else, tentatively [\ion{Xe}{3}] \lam5846.77.
The [\ion{Xe}{3}] \lam5846.77 identification, while not appearing in
the EMILI lists for NGC~7027 or IC~2501, is considered a better choice
than the other top ranked identification, [\ion{Fe}{2}] \lam5847.32,
which has poorer wavelength agreement and multiplet statistics in all
cases.  While the [\ion{Xe}{3}] $^3$P$_1$--$^1$D$_2$ transition at
1.37$\mu$ is unavailable for confirmation, we believe that the
[\ion{Xe}{3}] \lam5846.77 is definitely present in NGC~7027 and
IC~418, but is only tentatively identified in IC~2501.

The reality of the observed features potentially corresponding to the
[\ion{Xe}{4}] $^4$S$^{o}_{5/2,3/2}$--$^2$D$^{o}_{3/2}$
\lam\lam5709.2,7535.4 transitions is marginal except in NGC~7027.
However, their identifications as the Xe lines, should they be actual
emission lines, is more certain.  For \lam5709.2, the widely observed
\ion{N}{2} (V3) \lam5710.77 line is detected as a separate line in all
of the spectra.  Alternative identifications such as [\ion{Fe}{1}]
a5D$_3$--a5P$_1$ \lam5708.97 either have too many missing multiplet
members, except for IC~2501 where the ratio of potentially observed to
total number of multiplet lines expected is marginally better, or as
in the case of the \ion{Fe}{2}] \lam5709.04 intercombination line,
unlikely given that permitted \ion{Fe}{2} lines were not anywhere
definitively identified.  For \lam7535.2, the reality of the observed
lines is less uncertain, including IC~418 where a previously
unreported line has been uncovered by more thorough examination of its
spectra.  The \ion{Fe}{2}] (V87) \lam7534.82 identification proposed
by \citet{H01} is unlikely for the same reasons as \ion{Fe}{2}]
\lam5709.04 for [\ion{Xe}{4}] \lam5709.2.  The \ion{N}{2} 5f G
$^2$[7/2]$_4$--10d $^1$F$^{o}_{3}$ \lam7535.10 identification,
selected by \citet{E04} and \citet{P04}, also appears unlikely given
the relative strengths of other known permitted \ion{N}{2} lines in
the spectra.  It should be noted that the EMILI statistics for the
\lam7535.10 identifications in the MIKE PNe sample would probably have
been better if the line did not appear near the end of the last
spectral order where the wavelength calibration is the poorest.
Assuming an electric density well below the critical regime ($\sim
10^7$ cm$^{-3}$), the expected intensity ratio is
I(\lam5709.2)/I(\lam7535.4)=0.64 \citep{SB98}.  The observed range of
values that we observe, 0.33--1.19, suggests measurement errors on par
with the average values for such weak lines.  The strongest examples
in NGC~7027 do yield the best agreeing ratio (0.82).  As such, we
conclude that both [\ion{Xe}{4}] lines may be present in all PNe
spectra with varying degrees of certainty.

One of the stated goals for future spectroscopy of NGC~7027 and other
PNe given by PB94 was the detection of [\ion{Xe}{5}] transitions, none
of which they detected with any certainty.  In the present spectra, no
observed line passed the $\pm$1\AA\ initial selection criterion for
the $^3$P$_{1,2}$--$^1$D$_2$ \lam\lam5228.8,6998.7 lines, nor did
these transitions appear as possible EMILI IDs for any observed line.
This is despite the improved sensitivity and resolution which should
have enhanced the chances of detecting \lam5228.8, and which should
have easily separated \lam6998.7 from \ion{O}{1} (V21) \lam7002.10,
the contaminant noted by PB94.  Nevertheless, the detection of these
lines in our spectra would remain problematic since for \lam5228.8
extensive flaring from [\ion{O}{3}] \lam5006.84 in an adjacent order
leads to numerous ghosts in its vicinity, while for \lam6998.7
telluric absorption in the tail end of the Fraunhofer A band
complicates its observability.

EMILI did suggest that the fine-structure transition [\ion{Xe}{5}]
$^3$P$_0$--$^3$P$_2$ \lam7076.8 as a possible identification for
observed lines in two PNe, IC~4191 and NGC~7027, PNe with appropriate
excitation levels for the appearance of a line of this ion.  This is
the primary EMILI identification in IC~4191.  A third occurrence of
the corresponding observed line in NGC~2440 is clearly associated with
a nightglow line from the \Atm 3-2 band upon comparison with a
simulation of that band.  The competing identification \ion{C}{1}
(V26.01) \lam7076.48 has poor multiplet statistics, while
[\ion{Fe}{3}] \lam7078.10 and [\ion{Ni}{2}] \lam7078.04 are feasible,
but have poor wavelength agreement.  The \ion{Ca}{1} identifications
are also unlikely given that they arise from energy levels of large
energy, where lines that would follow from cascades from these levels
are not observed.  Two remaining obstacles to an identification with
[\ion{Xe}{5}] \lam7076.8 are the lack of the nebular $^3$P--$^1$D
lines and the low branching ratio for this electric quadrupole
transition ($\sim$0.008 with respect to $^3$P$_1$--$^3$P$_2$ at
2.07$\mu$).  Nevertheless, despite this transition's expected
weakness, the lack of viable alternate identifications suggests the
corresponding observed line may be at least tentatively identified in
NGC~7027, and probably identified in IC~4191, and both occurrences
yield reasonable abundance values in subsequent analysis.

The fine structure transition [\ion{Xe}{6}]
$^2$P$^{o}_{1/2}$--$^2$P$^{o}_{3/2}$ \lam6408.9 was identified with
certainty in NGC~7027 by PB94.  In the present NGC~7027 spectrum,
\lam6408.9 is probably associated with a weak but clearly separable
observed line on the red wing of \ion{He}{2} 5--15 \lam6406.38.  The
alternate primary EMILI identification of [\ion{Fe}{3}] \lam6408.50
was not considered likely, as an inspection of the NGC~7027 spectrum
for other lines from all low energy [\ion{Fe}{3}] multiplets did not
show a significant number of matches to warrant strong consideration
as a likely identification.  [\ion{Fe}{3}] \lam6408.50 also has a
comparatively high excitation energy (6.25 eV).  A \ion{C}{4} 9--17
\lam6408.70 identification is similarly downgraded by its high
excitation.  The reality of corresponding observed lines in NGC~2440
and IC~4191 is questionable, although for lack of suitable alternate
identifications, uncertain identifications of [\ion{Xe}{6}] \lam6408.90
are retained for both.

As with Kr, an inspection for auroral and trans-auroral transitions of
various Xe ions was also undertaken.  There were coincidences within
$\pm$1\AA\ between the [\ion{Xe}{3}] $^3$P$_1$--$^1$S$_0$ \lam3799.96,
[\ion{Xe}{3}] $^1$D$_2$--$^1$S$_0$ \lam5260.53, [\ion{Xe}{4}]
$^4$S$^{o}_{3/2}$--$^2$P$^{o}_{1/2}$ \lam3565.8, [\ion{Xe}{4}]
$^2$D$^{o}_{3/2}$--$^2$P$^{o}_{3/2}$ \lam4466.5, [\ion{Xe}{4}]
$^2$D$^{o}_{5/2}$--$^2$P$^{o}_{3/2}$ \lam5511.5, [\ion{Xe}{4}]
$^2$D$^{o}_{3/2}$--$^2$P$^{o}_{1/2}$ \lam6768.9, and [\ion{Xe}{5}]
$^1$D$^2$--$^1$S$_0$ \lam6225.3 lines.  However, except for two
instances, the identifications did not appear in the EMILI list for
those particular lines, and most had more reasonable and higher
ranked, identifications: \ion{C}{2} (V30) \lam5259.66,.76 for example
instead of the auroral [\ion{Xe}{4}] \lam5260.53 line.  In many cases
the line only appeared in one of the PNe spectra analyzed here, and
could be rejected due to the strength or absence of the nebular
transitions from the same ions.

\subsection{Br Line Identifications}
%6.3
In PB94 two transitions of Br were identified: [\ion{Br}{3}]
$^4$S$^{o}_{3/2}$--$^2$D$^{o}_{5/2}$ \lam6131.0 with certainty, and
[\ion{Br}{4}]$^2$D$^{o}_{3/2}$--$^2$P$^{o}_{3/2}$ \lam7385.1 as
possible.  A second nebular transition, [\ion{Br}{3}]
$^4$S$^{o}_{3/2}$--$^2$D$^{o}_{5/2}$ \lam6556.4, was lost in a blend
amidst the [\ion{N}{2}] \lam6548.04+H$\alpha$+[\ion{N}{2}] \lam6583.46
complex.

The source for the \ion{Br}{3} levels in the Atomic Line List v2.05 is
\citet{M58}, while the source used by PB94 is stated to be the
experimental levels listed in Table~IV of \citet{BH86a} from
unpublished work of Y.N. Joshi and Th.A.M. van Kleef, and provided
by private communication with van Kleef.  However, a comparison of the
Ritz-determined wavelengths from both sets of levels with those listed
by PB94 shows that the Joshi and van Kleef $^2$D$^o$ term energy
levels were used for the nebular transition wavelengths, while the
\citet{M58} levels were used for auroral transition wavelengths.  The
substantial difference in the $^2$D$^{o}_{3/2}$ level energy, 15042
cm$^{-1}$ for \citet{M58} and 15248 cm$^{-1}$ for Joshi and van Kleef,
leads to a difference in the $^4$S$^{o}_{3/2}$--$^2$D$^{o}_{3/2}$
transition air wavelength of 6646.3 \AA\ versus 6556.4 \AA, respectively,
with a lesser difference for the other transition, 6132.9 \AA\ versus
6131.0\AA.  Since the \citet{M58} levels used by the Atomic Line List
v2.05 and therefore by EMILI have stated uncertainties of 0.63
cm$^{-1}$, while those for Joshi and van Kleef in Table~IV of
\citet{BH86a} are listed to a precision 1 cm$^{-1}$, nearly the same
level of uncertainty, it is difficult to know which levels are more
accurate.  Therefore, EMILI was run using both sets of energy levels,
and the $^4$S$^o$--$^2$D$^o$ lines were searched for at both sets of
resultant wavelengths.

For the transitions wavelengths generated from the \citet{M58} energy
levels, \lam6132.9 and 6646.3, no observed line was detected in any
spectrum meeting the initial $\pm$1\AA\ selection criterion.  However,
for the $^4$S$^{o}_{3/2}$--$^2$D$^{o}_{5/2}$ transition wavelength
generated from those energy levels used by PB94, \lam6131.0, there do
appear to be observed lines in 4 of the 5 PNe spectra near the
\lam6130.4 line that PB94 identified with [\ion{Br}{3}].  As seen in
Table~\ref{tab9}, while [\ion{Br}{3}] \lam6131.0 is the highest
ranked EMILI identification in only one PN.  

However, other identifications for the putative [Br III] \lam6131.0
line are not compelling.  PB94 cite their observed line as being a
blend with \ion{C}{3} 7h $^{1,3}$H$^o$--16g $^{1,3}$G \lam6130.30,
which is the highest EMILI-ranked transition in each spectrum in which
the putative \ion{Br}{3} line is observed, except for IC~2501.
However, the companion line \ion{C}{3} 7h $^{1,3}$H$^o$--16i $^{1,3}$I
\lam6126.30, claimed by PB94 to have an equal intensity to
\lam6130.30, is not present in either the IC~4191 and IC~418 spectra.
Well-known lower excitation \ion{C}{3} lines are of negligible
intensity in IC~418, while in IC~2501 \ion{C}{3} \lam6126.30 has poor
wavelength agreement with its corresponding observed line, and is not
the primary EMILI identification for that line.  Only in NGC~7027 is
\ion{C}{3} \lam6126.30 a primary ID, suggesting that only in its
spectrum is the equally intense \ion{C}{3} \lam6130.30 likely to be
present and accountable for at least a portion of the putative
[\ion{Br}{3}] \lam6131.0 line.  As such for NGC~7027, the \lam6126.30
line intensity is subtracted from the observed line at 6130.32\AA,
while in the remaining spectra the corresponding observed lines are
ascribed wholly to [\ion{Br}{3}] \lam6131.0.  Possible contamination
due to flaring from an adjacent order in the NGC~7027 spectrum was not
accounted for.  The [\ion{Ni}{6}] \lam6130.40 identification, the
second ranked identification in many of the spectra, is of high
excitation (8.3 eV upper level), too high an ionization for IC~418,
and appears only to be ranked highly due to a favorable coincidence in
wavelength.

Spectroscopic confirmation for this identification is sought in the
possible presence of the [\ion{Br}{3}]
$^4$S$^{o}_{3/2}$--$^2$D$^{o}_{3/2}$ \lam6556.4 transition, in the PNe
spectra at expected intensities similar to that of \lam6131.0.  As
mentioned, detection of this line is difficult given its proximity to
the saturated H$\alpha$ line and its associated ghosts in its
immediate proximity.  Nevertheless, inspection of the line lists and
spectra (Figure~\ref{fig4}) does show distinct co-aligned features in
the IC~2501, IC~4191, and NGC~7027 spectra near this wavelength.
While initially assumed to be ghosts, the fact that these features can
be found in spectra from two different instruments and appear
invariant to the H$\alpha$ intensity, suggest they may be unrelated to
H$\alpha$.  That real lines can be observed between [\ion{N}{2}]
\lam6548 and H$\alpha$ is demonstrated by the presence of the strong
telluric OH P$_1$(3.5) 6-1 \lam6553.62 line just to the blue of the
suspected [\ion{Br}{3}] feature.  EMILI lists the [\ion{Br}{3}]
\lam6556.4 identification among possible choices in all PNe in which
the line appears, with the \lam6131.0 identification satisfying the
multiplet search in IC~2501 and IC~4191.  Alternate EMILI-favored
identifications for these features correspond to \ion{O}{2}, and
\ion{N}{2} permitted and core-excited transitions between levels of
primary quantum number 4--6, and are doubtful as they are of
comparatively high excitation with respect to better-known 3--3 or
3--4 transitions in these PNe that have equal or lesser intensities.
They appear here only due to their wavelength agreement with the
observed lines.  The highest ranked \ion{Fe}{2} \lam6555.94 is
discounted once again by the lack of other permitted Fe transitions in
any spectra.

The observed intensity ratios of the two putative [\ion{Br}{3}] lines,
I(\lam6131.0)/I(\lam6556.4), ranging from 0.18--0.45, do conflict with
theoretical expectation as follows.  While specific collision
strengths are unavailable at present for the \ion{Br}{3} states, the
ratio of the strengths relevant to these transitions should still be
roughly proportional to their respective upper level statistical
weights \citep{PB94,OF06}, weighted by a Boltzman factor respecting
the difference in level energies, and non-negligible fine-structure
emission between the $^2$D$^o$ levels, at sub-critical electron
densities.  Employing the IRAF \textit{nebular} package task
\textit{ionic} to solve for the relative populations of the $^2$D$^o$
levels at temperature derived from the diagnostic thought to be the
most appropriate, from [\ion{Ar}{3}], and using appropriate
\ion{Kr}{4} collision strengths as a proxy, scaled as discussed in
Section~\ref{sproabund}, the I(\lam6131.0)/I(\lam6556.4) ratio is
expected to be 0.72--0.83.  This is close to the 0.67 value expected
from the ratio of the collision strengths alone, and two to three
times larger than the ratio observed in the spectra.  At these levels
the [\ion{Br}{3}] \lam6131 line should have been observable in
NGC~2440, as the predicted intensity exceeds the likely detection
limit for features in its vicinity.  Therefore, we consider
[\ion{Br}{3}] to have been only tentatively detected in NGC~2440.
However, at least one [\ion{Br}{3}] line does appear to be present in
the remaining four PNe.  They all have lines detected at the same
wavelengths, although shifted 20--30 km s$^{-1}$ to the blue of the
wavelengths we have adopted for the [\ion{Br}{3}] lines, at
approximately the same relative intensities, and lack satisfying
alternate identifications.  The ratio of observed intensities (or in
the case of IC~418 the upper limit), while not exact, are within a
factor of 2--3 of those expected.

The identifications of the nebular [\ion{Br}{3}] lines suggests that
the level energies of \citet{M58} for at least the $^2$D$^o$ term
levels may be in error.  PB94 also identified an observed line at
7384.3\AA\ as possibly auroral $^2$D$^{o}_{3/2}$--$^2$P$^{o}_{3/2}$
\lam7385.2, but no such line appears in any of the PNe spectra
examined here.  However, since PB94 appears to have used
$^2$P$^o$--$^2$D$^o$ transition wavelengths derived from the
\citet{M58} energy levels, this might not be a surprise, although no
lines in any PN were discovered at the corresponding wavelength for
the $^2$D$^{o}_{3/2}$--$^2$P$^{o}_{3/2}$ transition (7483.1 \AA)
computed from the levels listed by \citet{BH86a} either.

The nebular [\ion{Br}{4}] $^3$P$_1$--$^1$D$^2$ \lam7368.0 line was
detected by PB94 at 7366.0\AA, affected by telluric absorption, and
blended with \ion{C}{4} 10--21 \lam7363.9 and \ion{O}{3} 4s
$^3$P$^{o}_{1}$--4p $^3$P$_1$ \lam7365.35.  In our NGC~7027 spectrum,
a line appearing at 7367.62\AA\ as a small protrusion above the local
continuum level interpolated between telluric absorption features in
both spectral orders covering that wavelength, is tentatively
identified as [\ion{Br}{4}] \lam7368.0.  The alternate candidates
listed by PB94, dielectronic \ion{C}{2} 3p$^\prime$
$^2$D$_{5/2}$--3d$^\prime$ $^2$P$^{o}_{3/2}$ \lam7377.00 is too far
away, while a \ion{O}{2} \lam7367.68 identification is doubtful given
its high upper level energy.  The \ion{C}{5} 7p $^3$P$^o$--8d $^3$D
\lam7367.60 transition is a possible alternate identification, but
only one other \ion{C}{5} transition in the spectrum, \ion{C}{5} 6gh
$^{1,3}$G,H$^o$--7hi $^{1,3}$H$^o$,I \lam4944.50 with an IDI value of
3, is a top ranked EMILI IDI; others are of lesser rank or do not
appear in the EMILI lists for the corresponding observed lines.

The other nebular transition, [\ion{Br}{4}] $^3$P$_2$--$^1$D$_2$
\lam9450.5 was not detected by PB94, even though the line should be of
the same intensity \citep{BH86a}.  Comparison between the IC~418 and
NGC~7027 spectra shows that something is filling in the telluric
absorption feature in the latter at 9450.85\AA, with an estimated
intensity close to [\ion{Br}{4}] \lam7368.10, although a ghost feature
at this wavelength arising from scattered light within the
spectrograph cannot be ruled out based on the proximity of other
similar features.  If the feature is real, however, the alternate
identification of \ion{Fe}{1} \lam9450.95 is not compelling.
Therefore, both transitions are identified in NGC~7027, although
tentatively since they appear in only one PN spectra, may be
attributable to misinterpretation of the local continuum level
complicated by telluric absorption of varying degree, and the latter
observed line may be spurious.  No auroral or trans-auroral features
of [\ion{Br}{4}] were identified in any spectrum.

\subsection{Other Z$>$30 Line Identifications}
%6.4
The spectra were searched for lines originating from other Z$>$30
ions.  Among the numerous coincidences between observed lines and
transition wavelengths at the $\pm$1\AA\ level, those listed in
Table~\ref{tab10} were judged to be the most likely to correspond to
real lines describable by a Z$>$30 identification in at least one of
the PNe.  Given the low abundances of these ions, only transitions
among the lowest-lying levels were expected to be observable.

PB94 declared as certain the identification of a line at 5758.7\AA\ in
NGC~7027 as [\ion{Rb}{4}] $^3$P$_2$--$^1$D$_2$ \lam5759.44.
Corresponding lines are detected in 4 of 5 PNe here.  However, at even
the highest instrumental resolution, this transition would blend with
\ion{He}{2} 5--47 \lam5759.74.  In NGC~2440 and IC~4191 there is
contamination from a flare or ghost at the position of \ion{He}{2}
5--47 \lam5759.74 in one spectral order.  Using only the
uncontaminated order, and inspecting the nearby \ion{He}{2} 5--n
sequence suggests that in IC~4191 the \ion{He}{2} \lam5759.74 line
shows significant excess, while in NGC~2440 there is better agreement
with recombination theory and the line is likely \ion{He}{2}.  The
NGC~7027 spectra, taken with a different instrument, does not show any
contamination, but does show a similar excess in \ion{He}{2} 5--47
\lam5759.74.  Thus, the line intensities in IC~4191 and NGC~7027 were
corrected by I(5--47)/I(5--46)=0.95 derived from the ratio of the
their emissivities \citep{SH95}.  IC~418 has no detectable \ion{He}{2}
lines, enhancing the [\ion{Rb}{4}] \lam5759.44 identification, but the
weakness and irregularity of its profile cast doubt upon its reality
as a line, and it is only tentatively identified here.  The alternate
EMILI recommended identification, [\ion{Fe}{2}] \lam5759.30, was not
considered likely given its high excitation energy.  Since the expected
intensity ratio of $^3$P$_2$--$^1$D$_2$ \lam5759.44 to
$^3$P$_1$--$^1$D$_2$ 9008.75 is 17 \citep{BH86b}, the apparition of
\lam9008.75 in IC~418 and NGC~7027 is probably likely due to
\ion{He}{1} 3d $^3$D--10p $^3$P$^o$ \lam9009.23,.26, or is unreal.  In
summary, the [\ion{Rb}{4}] \lam5759.55 line is identified in IC~4191,
NGC~7027, and tentatively in IC~418.

Another Rb line detection considered probable by PB94 in NGC~7027 is
[\ion{Rb}{5}] $^4$S$^{o}_{3/2}$--$^2$D$^{o}_{3/2}$ \lam5363.6, which
they identified at 5364.2\AA.  Although it does not appear in the
EMILI list for the corresponding lines in our spectra, we believe that
its identification in at least NGC~7027 is viable given the poor
multiplet statistics of [\ion{Ni}{4}] \lam5363.35, where none of the
four other multiplet lines are clearly present.  The \ion{O}{2} 4f F
$^2$[4]$^{o}_{7/2}$--4d$^\prime$ $^2$F$_{7/2}$ \lam5363.80
identification is an interesting alternative.  Well-known \ion{O}{2}
dielectronic doublet lines of multiplets V15, V16, and V36, such as
3s$^\prime$ $^2$D$_{5/2}$--3p$^\prime$ $^2$F$^{o}_{7/2}$ (V15)
\lam4590.97 are all clearly present in all of our PNe spectra, except
NGC~7027 where they either outside the bandpass or not optimally
placed for detection, at intensities $\sim 10^{-4}$ I(H$\beta$).
These lines are present at similar intensities in the NGC~7027
spectrum of \citet{Z05}.  Lines from 3d--4f transitions, particularly
3d $^2$D$_{5/2}$--4f F $^2$[4]$^{o}_{7/2}$ (V92a) \lam4609.4 arising
from the lower level of the \ion{O}{2} \lam5363.80 transition are also
favorably identified at roughly the same intensity.  Therefore it is
not out of the realm of possibility that this line is evidence of the
partial feeding of $^2$[4]$^{o}_{7/2}$ level.  However, while EMILI
did not perform a multiplet check on this line (due to a non
LS-coupled lower level), two other transitions at 5361.74\AA\ and
5375.57\AA\ from the same upper term and ending on this level are not
present in any of the spectra.  Without spontaneous emission
coefficients it is difficult to judge a potential branching ratio for
these transitions.  Therefore, it is believed that [\ion{Rb}{4}]
\lam5363.60 might be tentatively identified in IC~2501 and IC~4191 as
well.  Unfortunately the matching $^4$S$^{o}_{3/2}$--$^2$D$^{o}_{5/2}$
\lam4742.40 line is disguised by a flare or ghost in the MIKE spectra,
and is not optimally placed in NGC~7027 to allow a spectroscopic
confirmation, nor observed in IC~2501 and IC~4191.

Numerous lines of various Sr ions were searched for, including the
strong (I/I(H$\beta$)=$3\times10^{-4}$) [\ion{Sr}{4}]
$^2$P$^{o}_{3/2}$--$^2$P$^{o}_{1/2}$ \lam10276.9 fine-structure line
named as a possible identification by PB94 in NGC~7027, [\ion{Sr}{6}]
$^4$S$^{o}_{3/2}$--$^2$D$^{o}_{5/2}$ \lam4249.2 and
$^2$D$^{o}_{3/2}$--$^2$P$^{o}_{1/2}$ \lam5434.3 identified as possible
and tentative respectively by PB94 and also identified by \citet{Z04},
and the \ion{Sr}{2} resonance lines at \lam\lam4077.71,4215.52, and
the [\ion{Sr}{2}] $^2$S$_{1/2}$--$^2$D$_{5/2,3/2}$
\lam\lam6738.39,6868.17 lines observed in $\eta$ Carinae \citep{Z01}.
While, there were some instances of wavelength coincidences with
observed lines, the magnitude of the wavelength differences and
existence of plausible alternate identifications, such as
[\ion{Fe}{2}] 4249.08 for [\ion{Sr}{6}] \lam4249.2, did not warrant a
claim of identification.  The same was true for Y and Zr, where the
[\ion{Y}{5}] $^2$P$^{o}_{3/2}$--$^2$P$^{o}_{1/2}$ \lam8023.6 and
[\ion{Zr}{7}] $^3$P$_2$--$^3$P$_0$ \lam7961.4 fine-structure lines ,
listed as possible and tentative identifications by PB94 respectively,
did not have any wavelength coincidence with an observed line in any
PNe within the initial $\pm$1\AA\ screening criteria.
 
Turning to the fifth row of the periodic table, we believe that an
observed and previously unidentified line in IC~418 may correspond to
[\ion{Te}{3}] $^3$P$_1$--$^1$D$_2$ \lam7933.3, which would be the
first optical identification of a \ion{Te}{3} line in a PN.  The
\ion{He}{1} 3p $^3$P$^o$--ns $^3$S sequence of lines, of which the
alternate identifications \ion{He}{1} \lam\lam7932.36,.41 are members,
does not become clearly evident in this spectrum until 3p--10s at
8632.76\AA\ and 8632.83\AA.  The wavelength coverage of our spectra
does not extend out to the possibly stronger $^3$P$_2$--$^1$D$_2$
\lam10876.0 transition, so this line cannot be used to check the
\ion{Te}{3} identification.  Similarly, the detection of a possible
[\ion{I}{3}] $^4$S$^{o}_{3/2}$--$^2$D$^{o}_{3/2}$ \lam8536.6 line in
NGC~7027, with a less than compelling \ion{Cr}{2} permitted line as an
alternate primary EMILI identification, cannot be confirmed through
observation of its companion $^4$S$^{o}_{3/2}$--$^2$D$^{o}_{5/2}$
\lam6708.7 transition.  The putative \lam6708.7 line appears to be
better explained by a combination of [\ion{Mn}{2}] \lam6709.93 and
possibly [\ion{Cr}{5}] \lam6709.8, although the latter line has poor
multiplet statistics (2/0).  Nevertheless, the [\ion{I}{3}] \lam8536.6
transition has a definite IDI that is second ranked for its
corresponding observed line, so we count this as a tentative detection
for NGC~7027.
   
PB94 claims the detection of four transitions belonging to
permitted multiplets of \ion{Ba}{2}, with three
detections considered certain, and one considered possible.  Some of
these same transitions have also been identified in emission by
\citep{Z05} in NGC~7027 and by \citet{H04} in the compact H II region
within the Red Rectangle.  In the present sample it is believed that
one of the transitions considered certain by PB94, \ion{Ba}{2} 5d
$^2$D$_{5/2}$--6p $^2$P$^{o}_{3/2}$ \lam6141.71, may correspond to an
observed line in three of five PNe.  Given the low solar abundance and
high condensation temperature (1455 K; Lodders et al.\ 2003) of Ba,
the detection of these transitions might initially be considered
unlikely.  However, as originally proposed by PB94, if sufficient
gas-phase Ba$^+$ is available for collisional excitation, emission,
and moderate self absorption, it is estimated that optical depths of
0.5 and 2.7 in the 6s $^2$S$_{1/2}$--6p $^2$P$^{o}_{3/2}$ \lam4554.03
resonance transition are sufficient to account for both its apparent
non-detection, and for the observed intensity of the putative
\lam6141.71 lines in NGC 2440 and IC 4191 respectively.  PB94
estimated an optical depth of 3 for the \lam4554.03 transition in
NGC~7027.  Under these scenarios the \lam6141.71 line is either among
the strongest or is the strongest observable \ion{Ba}{2} line, with
its relative intensity with respect to other \ion{Ba}{2} lines
increasing with a larger optical depth in the \lam4554.03 line.

The EMILI results do suggest some potentially viable alternative
identifications other than \ion{Ba}{2} for individual PN, but none
that can satisfactorily account for the observed line in all three.
\ion{O}{1} 18d $^3$D$^{o}_{2}$--4p$^\prime$ $^3$D$_3$ \lam6141.75
arises from an auto-ionizing level, is too strong with respect to
other permitted \ion{O}{1} transitions further down the cascade chain,
and is not accompanied by any other multiplet members in any PNe.
Under the abundance and ionization model created by EMILI for NGC~2440
and IC~4191, \ion{Ne}{3} 4p $^5$P$_3$--4d $^5$D$^{o}_{4}$ \lam6141.48
did not produce an emission line with an intensity within three orders
of magnitude of the strongest predicted intensity among all putative
identifications.  This was not the case in NGC~7027, where the line
was among the strongest predicted lines, is expected to be the among
the strongest in the multiplet, and where the identification is
enhanced by favorable multiplet statistics.  The [\ion{Ni}{3}] 4s
$^3$F$_2$--4s $^3$P$_1$ \lam6141.83, arises from a level of high
excitation energy (9.9 eV), is probably not the strongest member of
its multiplet as it does not originate from the level of the multiplet
with the highest statistical weight, and other potentially stronger
multiplet members are not observed in IC~2501 and NGC~7027.  Yet in
NGC~2440, the multiplet statistics are somewhat better, and the large
intensity of the corresponding observed line and the Ba abundance
derived from it is at odds with lower abundances derived for Kr and Xe
in subsequent abundance analysis.  The excellent agreement
between the observed wavelengths in \textit{all} three PNe and the
\ion{Ba}{2} \lam6141.71 wavelength, the expectation that \lam6141.71
is the strongest \ion{Ba}{2} line, and the lack of a good alternate
identification for IC~4191, suggests that \ion{Ba}{2} \lam6141.71
warrants serious consideration as the correct identification in all
cases.

The fine-structure transition [\ion{Ba}{4}]
$^2$P$^{o}_{3/2}$--$^2$P$^{o}_{1/2}$ \lam5696.6, another certain
detection from PB94, is verified by its detection in our NGC~7027
spectrum after a correction for \ion{C}{3} (V2) \lam5695.92 is made.
The intensity attributable to \lam5696.6 was determined using the
effective recombination coefficient for \ion{C}{3} \lam5695.92
($\alpha=3.1\times10^{-15}$ cm$^3$ s$^{-1}$) specified by PB94, those
from \citet{NS84} and \citet{PPB91} for \ion{C}{3} 5g $^{1,3}$G--6h
$^{1,3}$H$^o$ \lam8196.61, and the \lam8196.61 line's observed
intensity.  Subtraction yielded a line amounting to 36\% of the
originally observed intensity.

The only sixth-row elemental transition sought in these spectra was
the fine-structure [\ion{Pb}{2}] $^2$P$^{o}_{1/2}$--$^2$P$^{o}_{3/2}$
\lam7099.80 line, which was identified with certainty by PB94.  The
transition appears in the EMILI list for IC~418 tied for the highest
ranked transition for a previously unidentified line.  For NGC~7027,
the transition is not ranked, but the competing identifications in its
EMILI list are not convincing given their high excitation energies.
Excellent wavelength agreement is seen in both cases, and both
identifications are considered certain.

In summary, of the 18 Z$>$30 elemental ion transitions considered
certain or probable by PB94 in NGC~7027, 15 are believed to be
detected to various degrees of certainty in the present set of
spectra.  The final intensities of all Z$>$30 lines detected with any
certainty within at least one PN are presented in Table~\ref{tab11},
with certain identification listed in bold, and tentative
identifications in normal type.  Conspicuous among those missing from
PB94 is [\ion{Se}{3}] $^3$P$_1$--$^1$D$_2$ \lam8854.2.  The nearby
\ion{He}{1} 3d $^3$D--11p $^3$P$^o$ \lam8854.14 transition does appear
to show a large excess relative to that expected from other confirmed
lines in the same series in both IC~418 and IC~7027.  However, the
energy levels in the Atomic Line List v2.05, derived from \citet{M52}
and also utilized by PB94, have a listed uncertainty of 6.3 cm$^{-1}$
that exceeds the maximum 1 cm$^{-1}$ tolerance allowed for inclusion
of supplemental Z$>$36 ions into EMILI.  This leads to an error of up
to 10\AA\ in the transition wavelength.  Thus, the identification
cannot be confirmed under the present degree of energy level
uncertainty.  Also missing are the \ion{Ba}{2} transitions 6s
$^2$S$_{1/2}$--6p $^2$P$^{o}_{3/2}$ \lam4554.03 and 5d
$^2$D$_{3/2}$--6p $^2$P$^{o}_{3/2,1/2}$ \lam\lam5853.67,6496.90.  As
discussed previously, all would be weaker in our spectra than the
potentially observed \lam6141.71 under the assumption of a moderate
optical depth in the resonance transition \lam4554.03 (and 6s
$^2$S$_{1/2}$--6p $^2$P$^{o}_{1/2}$ \lam4934.08 not observed by PB94).

\citet{Z05} has claimed the identification of 5 additional transitions
in their NGC~7027 spectra belonging to Z$>$30 elements that were not
observed in our spectra.  Two additional permitted \ion{Ba}{2} lines,
\ion{Ba}{2} \lam4554.03 and \ion{Ba}{2} \lam4934.08 are detected, both
weaker than the \lam6141.71 line as would be expected for substantial
optical depths of the resonance lines.  They also identify auroral
[\ion{Br}{3}] \lam7385.2 and [\ion{Rb}{5}] \lam5080.2, although the
nebular [\ion{Br}{3}] \lam6131.0 and [\ion{Rb}{5}]
\lam\lam4742.4,5363.6 lines, which might be expected to be stronger
than the auroral lines, are not identified.  \citet{Z05} also
identifies [\ion{Sr}{6}] \lam4249.2, which might be better
attributable to [\ion{Fe}{2}] \lam4249.08.

\subsection{Z$>$30 Line Identifications in \ion{H}{2} Regions}
%6.5
In comparing the spectra of the PNe with \ion{H}{2} regions, we note
that no post-Fe peak lines were identified by \citet{GR04} in their
deep VLT UVES echelle spectrum of the \ion{H}{2} region NGC~3576.  In
Table~\ref{tab11} are included estimates of the upper limits to
undetected Kr and Xe line intensities for this \ion{H}{2} region from
the faintest detectable lines that were identified at neighboring
wavelengths or from artificial lines inserted at their wavelengths
that meet minimal detection S/N statistics.  In the spectrum of the
Orion Nebula, \citet{B00} reported weak (I$\approx 2\times 10^{-5}$
H$\beta$) unidentified features at 5867.8 \AA\ and 6826.9 \AA\ that
were originally unidentified, but can now be positively identified as
[\ion{Kr}{4}] $^4$S$^{o}_{3/2}$--$^2$D$^{o}_{3/2}$ \lam5867.74 and
[\ion{Kr}{3}] $^3$P$_2$--$^1$D$_2$ \lam6826.70 respectively.  However,
there are no other close coincidences between any lines listed in
Tables~\ref{tab7}--\ref{tab10}, except for 5363.34\AA\ which would
correspond with [\ion{Rb}{5}] $^4$S$^{o}_{3/2}$--$^2$D$^{o}_{3/2}$
\lam5363.60, an unlikely identification given the unrealistically high
degree of ionization (40 eV) for an H II region.  The observed
intensities and upper limits are also included in Table~\ref{tab11}.

\section{Z$>$30 Abundances} \label{sproabund}
%7
The line intensities given in Table~\ref{tab11} were used to compute
abundances for Ba, Kr, Xe, and Br ions.  The atomic data listed in
Table~\ref{tab3} were formatted for inclusion in 5-level atom models
for abundance analysis with the IRAF \textit{nebular} task
\textit{abundance}, or for the cases of the fine structure lines
[\ion{Xe}{6}] \lam6408.9 and [\ion{Ba}{4}] \lam5696.6, using a
two-level atomic solution code.  For the [\ion{Br}{3}] and
[\ion{Br}{4}] lines the relevant collision strengths have not yet been
calculated.  However, since these ions are isoelectronic with
[\ion{Kr}{4}] and [\ion{Kr}{5}], and because collision strengths for
the same levels along an isoelectronic sequence tend to vary with
effective nuclear charge \citep{S58}, with some exceptions, the
collision strengths of [\ion{Br}{3}] and [\ion{Br}{4}] were assumed to
be 25\% smaller than those for Kr.  Because collision strengths were
not calculated by \citet{SB98} for [\ion{Xe}{5}] transitions, the
[\ion{Kr}{5}] collision strengths for the same transitions were
utilized in the Xe$^{+4}$ analysis.  Collision strengths for
excitation to the $^3$P$_2$ parent level of the tentatively observed
[\ion{Xe}{5}] 5p$^2$ $^3$P$_0$--$^3$P$_2$ \lam7076.8 line in other
np$^2$ ions such as \ion{Ne}{5} \citep{LB94}, \ion{Ar}{5}
\citep{GMZ95}, and \ion{Kr}{5} \citep{Sc97}, depart by 15\% at most
from their \ion{Kr}{5} values at 10000 K.  The temperatures and
densities used for these analyses corresponded to those from
diagnostics with the closest ionization potential.  The [\ion{Ar}{3}]
(27.6 eV) temperature and [\ion{Cl}{3}] (23.8 eV) density diagnostic
values were used for [\ion{Br}{3}] (21.8 eV), [\ion{Kr}{3}] (24.4 eV),
[\ion{Xe}{3}] (21.1 eV), and [\ion{Ba}{4}] (20.0 eV); [\ion{O}{3}]
(35.1 eV) temperature and [\ion{Ar}{4}] (40.1 eV) density for
[\ion{Br}{4}] (36.0 eV), [\ion{Kr}{4}] (37.0 eV), and [\ion{Xe}{4}]
(32.1 eV); [\ion{Ne}{3}] (41.0 eV) temperature [\ion{Ar}{4}] (40.1 eV)
density for [\ion{Xe}{4}] (46.0 eV); [\ion{Ar}{5}] (59.8 eV)
temperature and [\ion{K}{5}] (60.9 eV) density for [\ion{Kr}{5} (52.5
eV)] and [\ion{Xe}{6}] (57.0 eV).  The [\ion{O}{1}] temperature and
[\ion{N}{1}] density were used for \ion{Ba}{2} assuming collision
excitation of Ba$^+$ as the source of \ion{Ba}{2} \lam6141.71.
Averaged diagnostic values were used when one of the diagnostics was
unavailable for a particular ion.  Uncertainties were computed in the
same manner as the lighter elements, through permutation of intensity
measurement and diagnostic value errors, where available, and
selection of extrema values.  For lines that were corrected for
blending, 25\% of the value of the correction was added to the
measurement uncertainty.

Table~\ref{tab12} presents the results of the abundance determinations
for individual lines of Ba, Br, Kr, and Xe ions.  To compute overall
elemental abundances for the latter three ions, argon has been
selected as a benchmark element in addition to hydrogen.  This was
done because the ionization potentials of the noble gases Ar, Kr, and
Xe, as well as Br, are very similar for their first three stages of
ionization and therefore ionization corrections should not be large
when making abundance comparisons among these elements.  Additionally,
the noble gases are almost completely non-reactive and have very low
condensation temperatures, so corrections for gas phase abundances
depleted by grain formation are insignificant for these elements.
Only ionic abundances relative to hydrogen were computed for Ba.

We convert from ionic to elemental abundances by making use of the
similarity in ionization potentials of the noble gases, so that
\begin{eqnarray*}
\textrm{Kr}/\textrm{Ar} & = & (\textrm{Kr}^{+2} +
\textrm{Kr}^{+3})/(\textrm{Ar}^{+2}+\textrm{Ar}^{+3})\, {\rm and} \, 
(\textrm{Kr}^{+2}+\textrm{Kr}^{+3}+\textrm{Kr}^{+4})/(\textrm{Ar}^{+2}+\textrm{Ar}^{+3}+\textrm{Ar}^{+4})
\,, \\
\textrm{Xe}/\textrm{Ar} & = & (\textrm{Xe}^{+2} +
\textrm{Xe}^{+3})/(\textrm{Ar}^{+2}+\textrm{Ar}^{+3})\, {\rm and} \, 
(\textrm{Xe}^{+2}+\textrm{Xe}^{+3}+\textrm{Xe}^{+4}+\textrm{Xe}^{+5})/(\textrm{Ar}^{+2}+\textrm{Ar}^{+3}+\textrm{Ar}^{+4})
\,,
\end{eqnarray*}
for H II regions and PNe judged to be of low excitation (IC~2501 and
IC~418), and for high-excitation PNe (NGC~2440, IC~4191, NGC~7027),
respectively.  The Kr$^{+4}$/H$^+$ abundance was included in the
IC~2501 Kr/Ar ratio determination.  For NGC~7027, the Br/Ar ratio was calculated
the same way as the Xe/Ar ratio, with corrections for unobservable
Br$^{+4}$ and Br$^{+5}$ made assuming
Br$^{+4}$,Br$^{+5}$/Br=Xe$^{+4}$,Xe$^{+5}$/Xe, appropriate because of
their very similar ionization potentials.

The resulting abundances are listed in the bottom rows of
Table~\ref{tab12} relative to their solar values from \citet{L03}.
Three particular results are noteworthy.  First, three out of five PNe
show significant enhancements in Kr and Xe relative to solar values,
while two others show similar, solar-like values.  It is also
interesting to note that IC~418 and NGC~7027, both considered young
PNe, have the largest over-abundances, although they are of greatly
different ionization classes.  Meanwhile, the H II regions show only
solar Kr abundances, indicative of unprocessed ISM gas.  Secondly, the
Kr and Xe abundances in all PNe all show enhancements of similar
magnitude, as was seen in NGC~7027 by PB94.  Finally, for NGC~7027,
the Br abundance also shows a level of enhancement similar to that of
Kr and Xe.  Uncertainties remain regarding the adaptability of Kr
collision strengths, modified as was done here, for the Br abundance
calculations.  However, to reduce the Br abundance to solar, the
collisions strengths for Br$^{+2}$ and Br$^{+3}$ would have to both be
four times greater than their counterparts in Kr, and this is much
larger than the difference seen for analogous transitions for other
p$^2$ and p$^3$ ion pairs (N and O, Cl and Ar).  In any event, the
prevalence of probable Br line identifications suggests, independent
of the actual atomic data, that significant Br does exists in most of the
PN comprising our sample.

It is of interest to compare the s-process abundances we derive for
PNe with those obtained for evolved stars from analyses of their
absorption spectra.  With the exception of Rb, Ba, and Pb, the
elements observed in emission in PNe are different from those normally
observed in the spectra of late-type stars, so a direct comparison is
not feasible.  In fact, lines of the same $s$-process elements are not
necessarily even observed in similar-type stars.  For this reason
\citet{LB91} defined two parameters, [ls] and [hs], that represent the
mean abundance of elements associated with Sr and Ba, respectively, in
what are called the ``light'' and ``heavy'' $s$-process peaks.  They
define the abundance indices [ls] and [hs] for an object as the mean
logarithmic abundances relative to iron of (Y, Zr, \& Sr) and (Ba, Nd,
La, \& Sm), respectively, compared to their solar mean abundances.
All of these elements are produced by the $s$-process, although
several of them have predominantly $r$-process contributions for
solar-type stars \citep{A99}.

Of the post-Fe peak elements that are observed in PNe, Kr belongs to
the light $s$-process peak near Sr, and Xe is near Ba in the heavy
$s$-process peak.  Both are produced by the $s$-process, although Xe
is predominantly an $r$-process element for stars of solar metalicity.
Using models to determine the appropriate correction factors, Kr and
Xe can be incorporated into the [ls] and [hs] indices.  However, in
making comparisons of nebular abundances with those derived from
stellar spectra Fe should not be used as the fiducial abundance for
nebulae because its consistently strong depletion from the gas phase
due to grain formation causes its true abundance in nebulae to be
indeterminate \citep{S75,P99}.  Argon is a good surrogate for Fe
because it does not suffer depletion in nebulae, and it is well
observed in nebulae in multiple ionization stages and the relevant
excitation cross sections are known.

The extent to which $s$-process nucleosynthesis occurs in stars is
determined by the total neutron exposure after the Third Dredge Up
phase.  Calculations show that neutron exposure is the primary factor
that determines the ratio of the enhancements of the heavy to light
$s$-process elements, i.e., [hs/ls], the
``neutron-exposure-related-parameter''.  Low neutron exposures result
in element production confined to a crowded region around the Sr peak,
producing [hs/ls]$<0$, whereas higher neutron exposures significantly
populate the Ba peak, leading to [hs/ls]$>$0 \citep{Bu95,BGW99}.  The
fact that [Xe/Kr]$\approx$1 for PNe, in spite of Xe not being produced
by the s-process as much as Kr, is strongly suggestive that the
progenitor AGB stars of PNe experience significant neutron
exposure. Of equal significance, the fact that forbidden lines of
post-Fe peak elements not observed in stars are detectable in nebulae
causes \ion{H}{2} regions and PNe to be of potentially great value in
studying the nucleosynthesis of these elements.

\section{Summary}
%8 
In summary, very high signal-to-noise spectra of PNe do reveal lines
from elements that are enhanced by the $s$-process, as was also found
by other investigators.  Because many post-Fe peak elements are
refractory their gas phase abundances are not reliable indicators of
the nucleosynthetic processes that have occurred in the progenitor
stars.  Fortunately, the noble gases Ar, Kr, and Xe are largely
unaffected by molecular and grain formation, and the latter two
elements are situated in the light and heavy $s$-process peaks,
respectively.  They also have easily excitable and observable lines in
multiple ionization stages for which atomic cross sections are now
available, and whose analyses should be straightforward.

We find for a sample of five PNe that Kr and Xe abundances are
enhanced over solar values by up to an order of magnitude, from both
an analysis of their intensities relative to those of the fiducial
element Ar, and also from a relative comparison of their line
strengths in PNe vs.  \ion{H}{2} regions such as the Orion Nebula,
where the lines are very weak in gas that is representative of the ISM
composition.  The similar enhancements of Kr and Xe in PNe are
suggestive of large neutron exposures in the progenitor central stars.
Further spectroscopy of PNe should reveal additional post-Fe peak
emission lines whose analyses will contribute to a more complete
picture of post-main sequence stellar evolution, while deep spectra of
\ion{H}{2} regions should lead to improved values of the ISM
abundances of these elements

\acknowledgments The work of YZ and XWL was partially supported by
Chinese NSFC Grant NO.10325312, and YZ gratefully acknowledges the
award of an Institute Fellowship from STScI, where his work was
carried out.  EP, KC, and JAB gratefully acknowledge support for this
work from NSF grant AST-0305833 and HST grant GO09736.02-A.

\clearpage
\input{tab1}

\clearpage
\input{tab2}
\clearpage
\input{tab3}
\clearpage
\input{tab4}
\clearpage
\input{tab5}
\clearpage
\input{tab6}

\clearpage
\pagestyle{empty}
\input{tab7}
\clearpage
\input{tab8}
\clearpage
\input{tab9}

\clearpage
\input{tab10}

\clearpage
\pagestyle{plaintop}
\input{tab11}

\clearpage
\input{tab12}
\clearpage

\clearpage

\clearpage

\clearpage

\clearpage

\clearpage
{\plotone{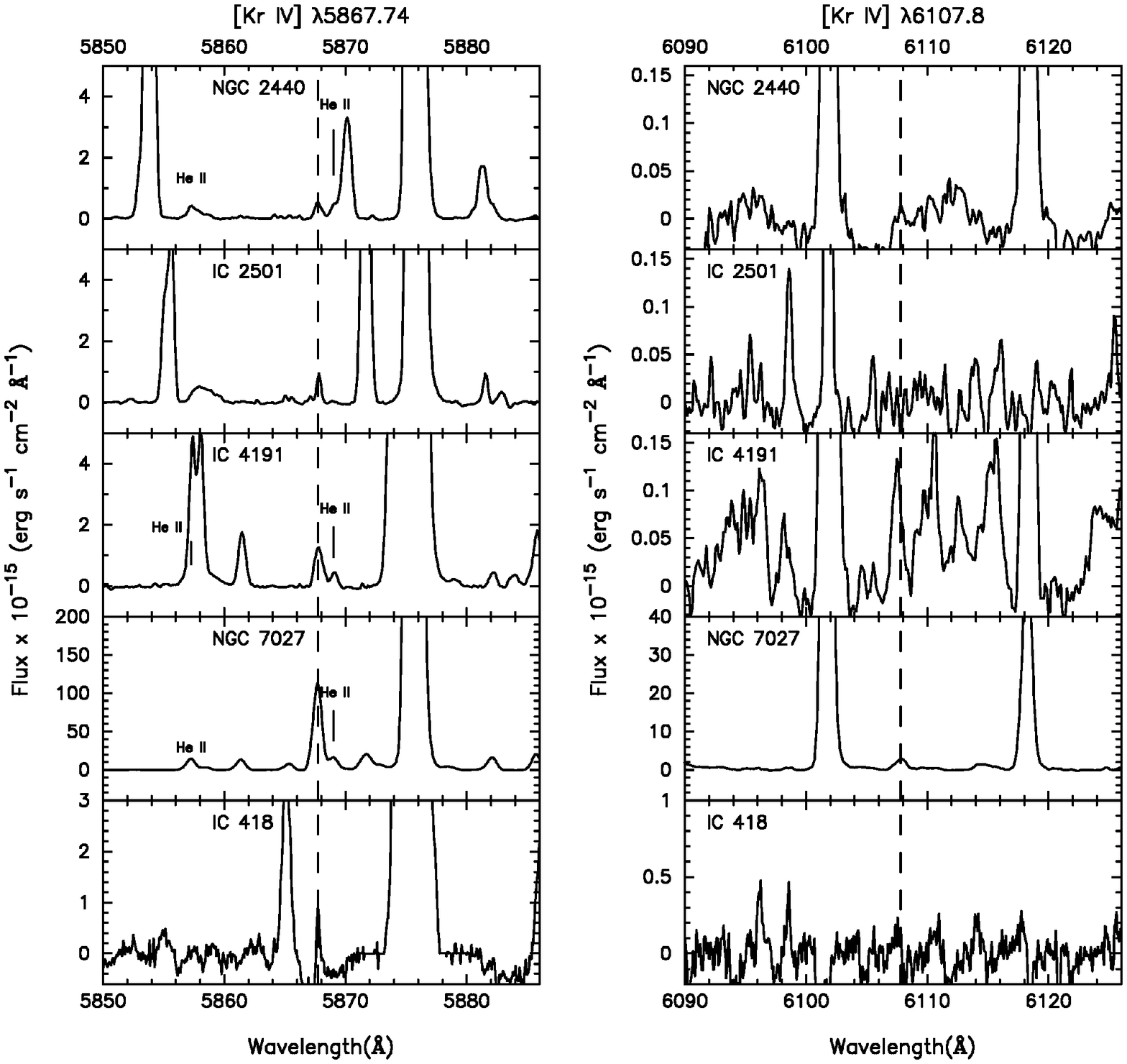}}\\[5mm]
\centerline{Fig. 4. --- Continued.}
\clearpage
{\plotone{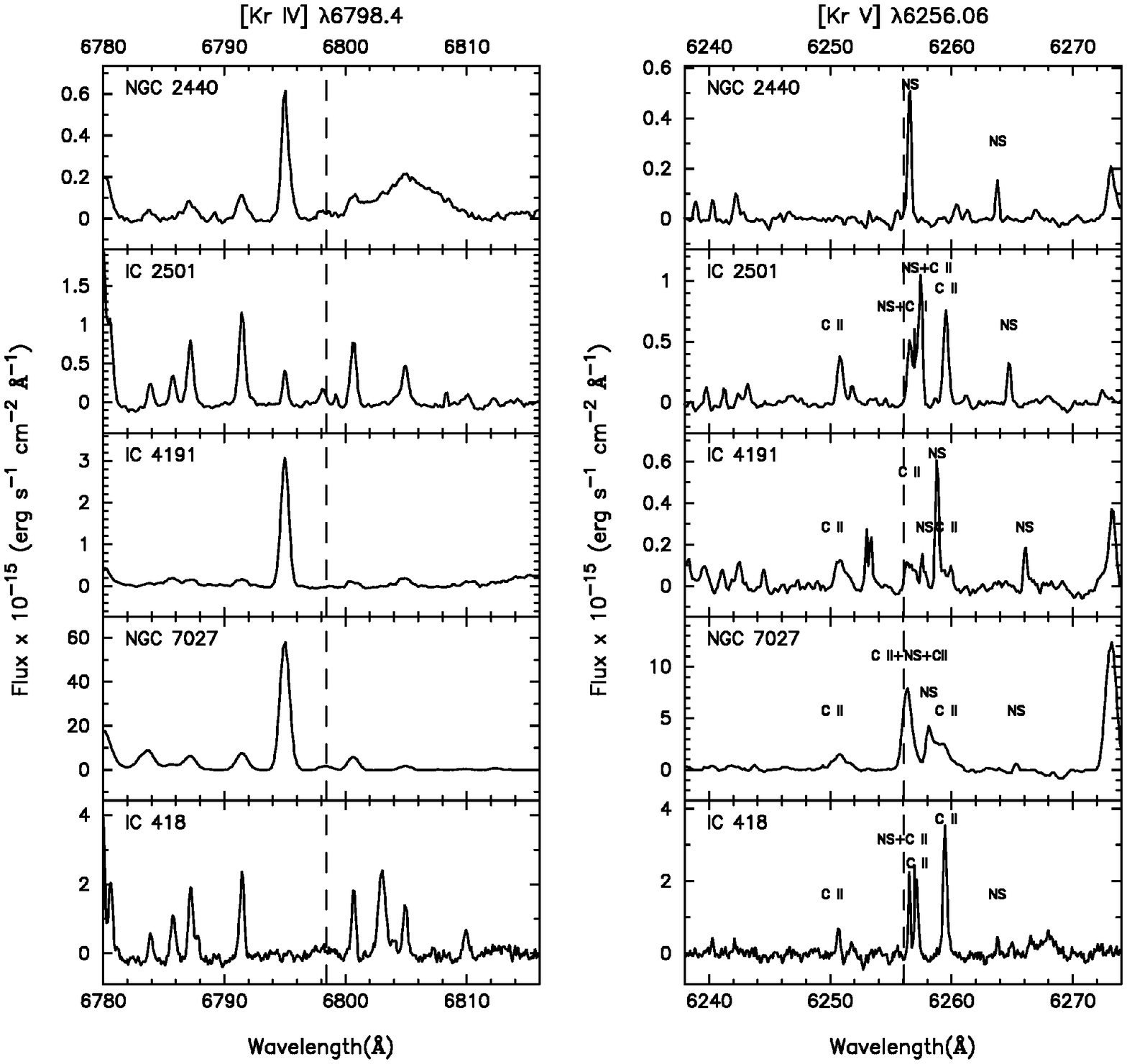}}\\[5mm]
\centerline{Fig. 4. --- Continued.}
\clearpage
{\plotone{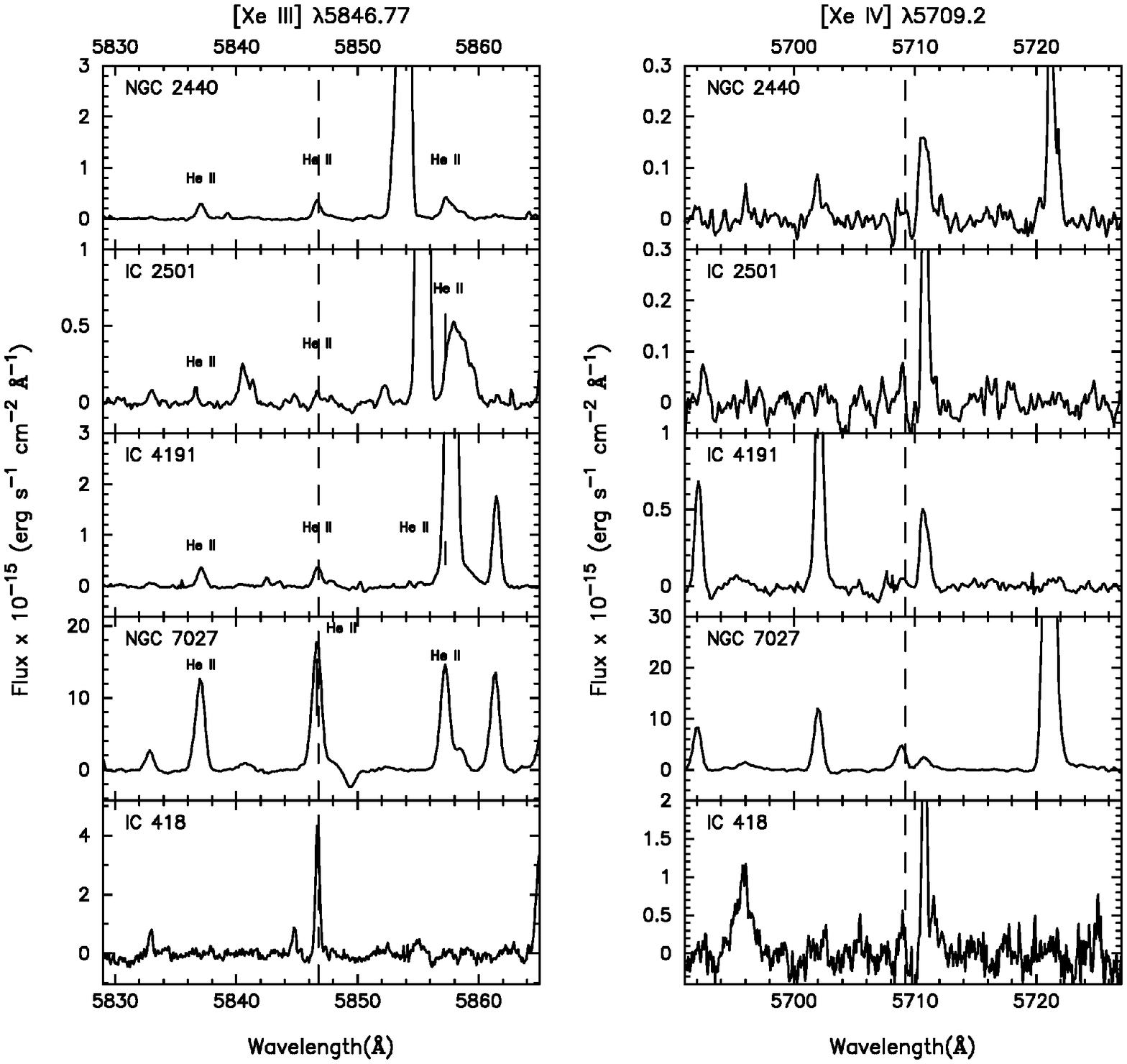}}\\[5mm]
\centerline{Fig. 4. --- Continued.}
\clearpage
{\plotone{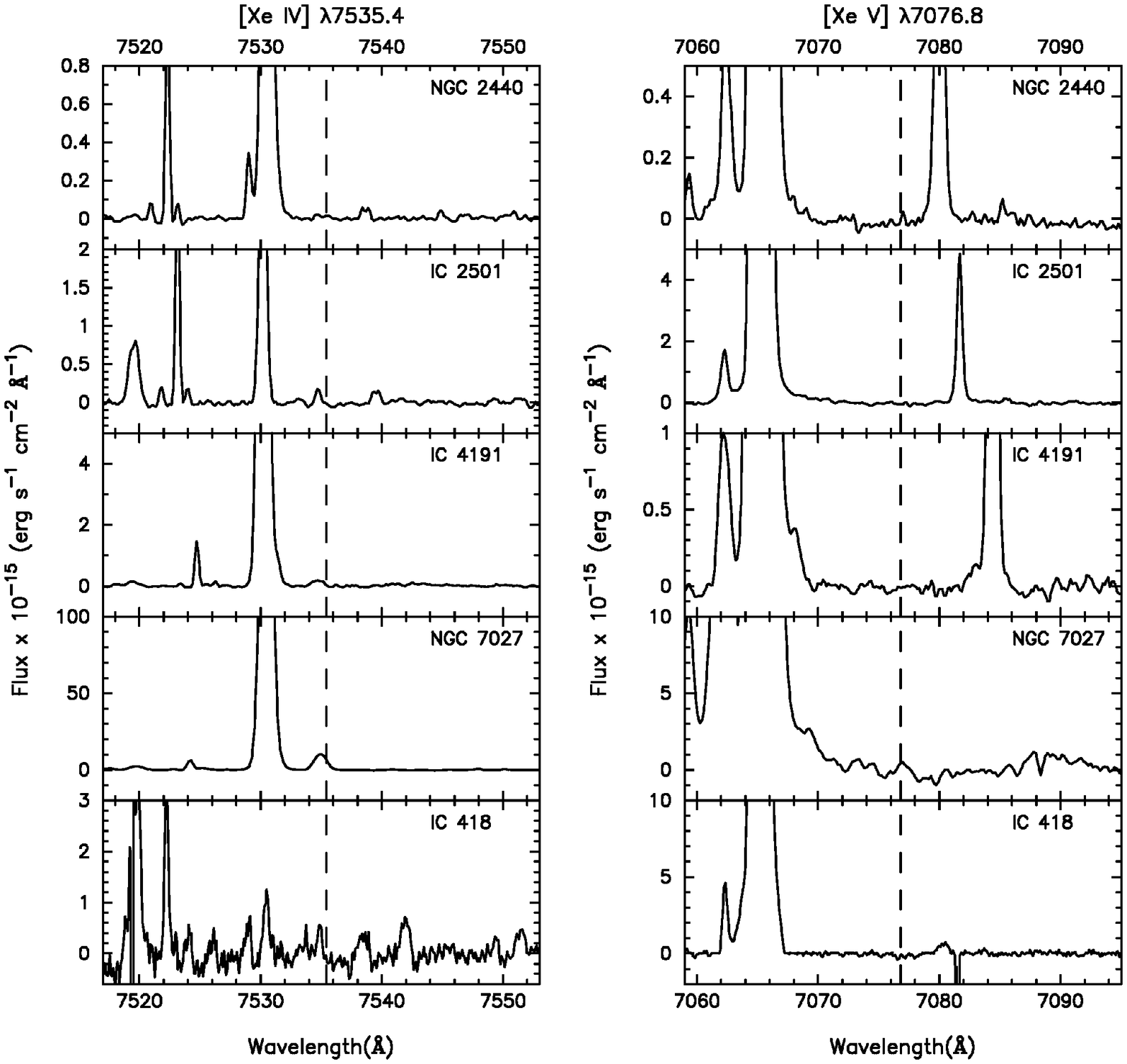}}\\[5mm]
\centerline{Fig. 4. --- Continued.}
\clearpage
{\plotone{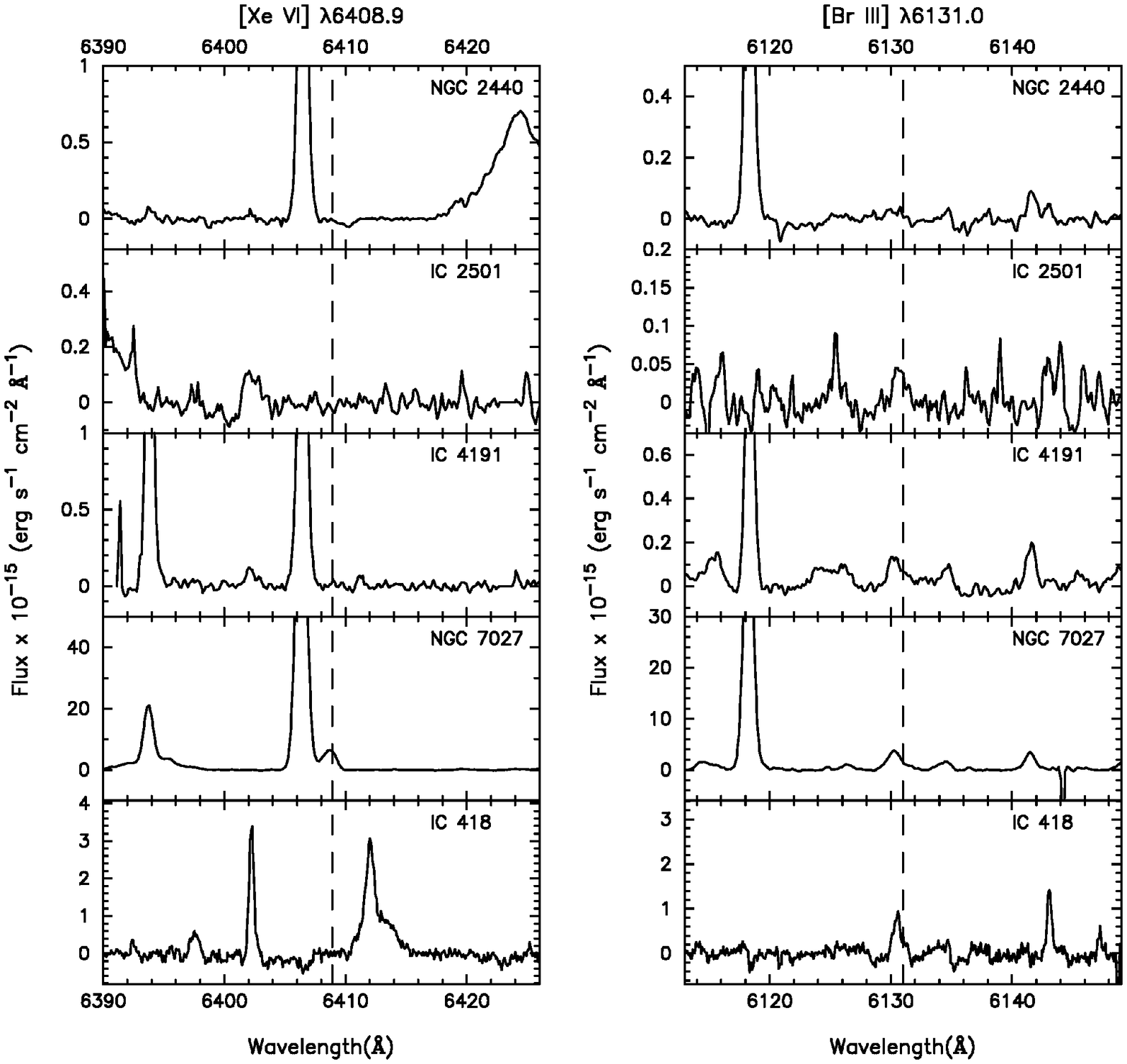}}\\[5mm]
\centerline{Fig. 4. --- Continued.}
\clearpage
{\plotone{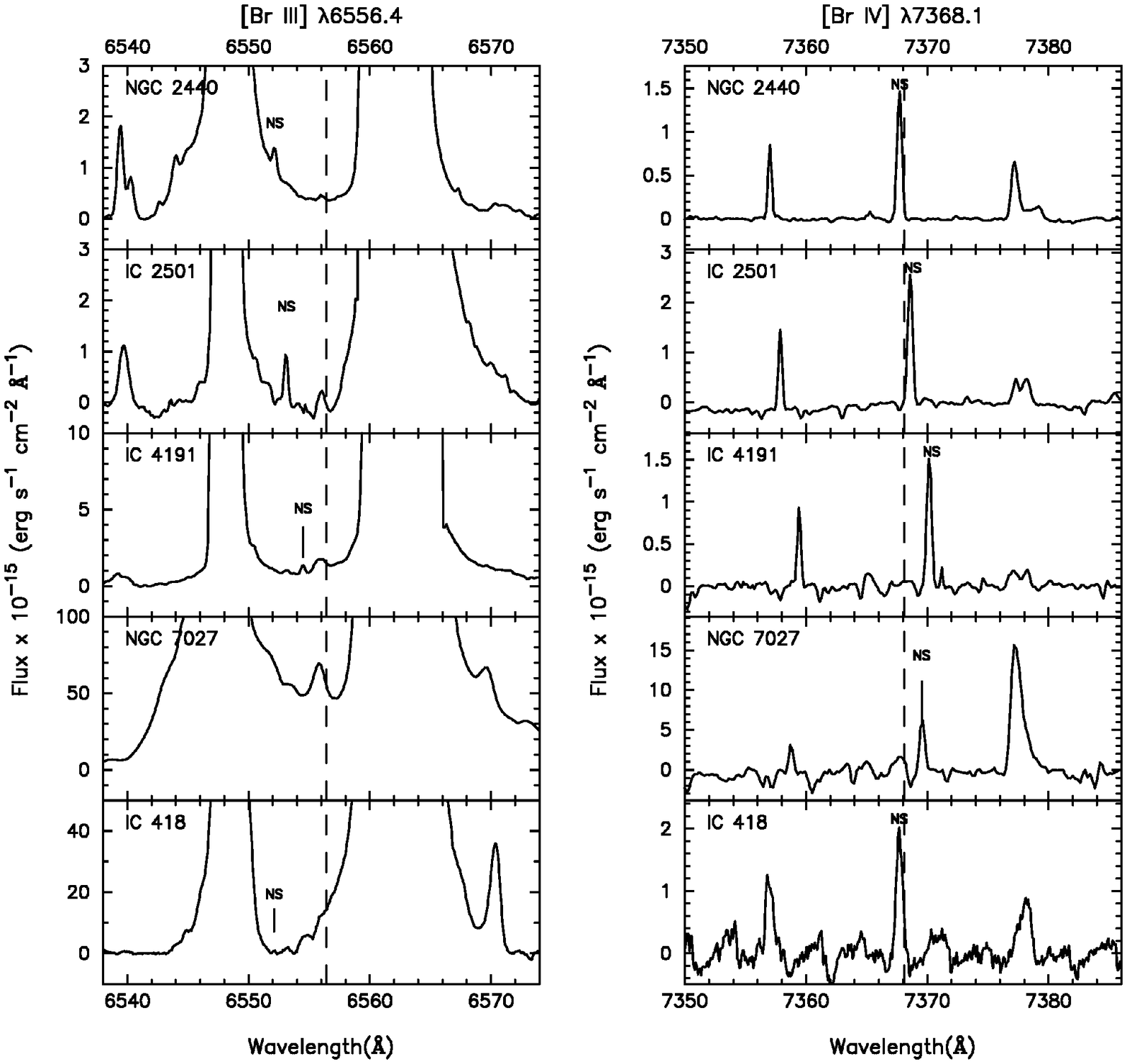}}\\[5mm]
\centerline{Fig. 4. --- Continued.}

\clearpage

\clearpage

\end{document}

%% file: tab1.tex
\begin{deluxetable}{lllll}
\setlength{\tabcolsep}{0.05in}
\tablecaption{Journal of Observations
\label{tab1}}
\tablewidth{0pt}
\tablehead{
\colhead{Object}& \colhead{Telescope} &
\colhead{Dates(UT)} & \colhead{Integration Times(sec) } & \colhead{Slit Center}
}
\startdata
NGC~7027&KPNO 4m  & 19--22 Jun 2002 &34$\times$1200s, 9$\times$300s, 1$\times$120s, & 3.5$''$ W, 0.5$''$ N \\
&  &  &4$\times$60s, 6$\times$30s   & \\
IC~2501 &LCO 6.5m & 12--13 Feb 2003 &6$\times$1800s, 1$\times$300s, 4$\times$60s & 2$''$ E\\
IC~2501 &LCO 6.5m & 25--26 Feb 2003 &15$\times$1800s, 2$\times$60s & 2$''$ E\\
IC~4191 &LCO 6.5m & 12--13 Feb 2003 &5$\times$1800s, 1$\times$1200s, 1$\times$300s,1$\times$60s & 2$''$ E\\
IC~4191 &LCO 6.5m & 25--26 Feb 2003 &10$\times$1800s,1$\times$300s,3$\times$60s & 2$''$ E\\
NGC~2440&LCO 6.5m & 12--13 Feb 2003 &11$\times$1800s, 1$\times$300s, 1$\times$60s &14$''$ E\\
\enddata
\end{deluxetable}

%% file: tab2.tex
\begin{deluxetable}{lcccc}
\tablewidth{0pt}
\tablecaption{Nebular Parameters \label{tab2}}
\tablehead{
\colhead{Parameter} &
\colhead{NGC~2440} & \colhead{IC~2501} & \colhead{IC~4191} &
\colhead{NGC~7027}
}
\startdata
V(geo) (km s$^{-1}$) & +66.7 & +23.2 & $-$40.1 & $-$7.1 \\
c(H$\beta$)          &  0.55 &  0.55 &    0.77 &   1.17 \\
\enddata
\end{deluxetable}

%% file: tab3.tex
\begin{deluxetable}{lrrrr}
\tablecaption{Electron Temperatures and Densities \label{tab3}}
\tablewidth{6.25in}
\tabletypesize{\footnotesize}
\tablehead{
\multicolumn{1}{c}{Diagnostic} & 
\multicolumn{1}{c}{NGC~2440} &
\multicolumn{1}{c}{IC~2501} &
\multicolumn{1}{c}{IC~4191} &
\multicolumn{1}{c}{NGC~7027} 
}
\startdata
\multicolumn{5}{l}{Density (cm$^{-3}$)} \\
\textrm{[}N I\textrm{]} \lam5198/\lam5200    &  3300$^{+2700}_{-1300}$ & 21000$^{.....}_{-15000}$ & 12000$^{.....}_{-7000}$ & 15000$^{.....}_{-8000}$ \\
\textrm{[}S II\textrm{]} \lam6716/\lam6731   &  3100$^{+1200}_{-700}$  & 11000                    & \nodata                 & \nodata                 \\
\textrm{[}O II\textrm{]} \lam3726/\lam3729\tablenotemark{a}  &  3300$^{+1100}_{-800}$  & 11000$^{+9000}_{-4000}$  &  9000$^{+6000}_{-3000}$ & \nodata\\
              \hspace{6.8pc}               \tablenotemark{b}  &  3700                   & 11000                    &  \nodata                & \nodata\\
\textrm{[}Cl III\textrm{]} \lam5517/\lam5537 &  4700$^{+1100}_{-900}$  & 8500$^{+2100}_{-1600}$   & 12000$^{+3000}_{-2000}$ & 47400$^{+1900}_{-1800}$ \\
\textrm{[}Ar IV\textrm{]} \lam4711/\lam4741\tablenotemark{c} &  6200$^{+1600}_{-1300}$ & 9300$^{+2100}_{-1600}$   & 11800$^{+2300}_{-1900}$ & \nodata\\
                              \hspace{7.6pc}\tablenotemark{d} &  5900$^{+1600}_{-1300}$ & 7400$^{+2200}_{-1500}$   & 11000$^{+2400}_{-1800}$ & 49200$^{+1300}_{-1200}$ \\
\enddata
\tablenotetext{a}{Versus [O II] temperature diagnostic.}
\tablenotetext{b}{Versus [N II] temperature diagnostic.}
\tablenotetext{c}{Versus [O III] temperature diagnostic.}
\tablenotetext{d}{Versus [Cl IV] temperature diagnostic.}
\tablenotetext{e}{Versus [Ne III] temperature diagnostic.}
\end{deluxetable}

%% file: tab4.tex
\begin{deluxetable}{lrrrr}
\tablecaption{Ionic Abundances \tablenotemark{a} \label{tab4}} 
\tablewidth{4in}
\tabletypesize{\footnotesize}
\tablehead{
\multicolumn{1}{c}{\textit{X}$^{+i}$/H$^+$} &
\multicolumn{1}{c}{NGC~2440} &
\multicolumn{1}{c}{IC~2501} &
\multicolumn{1}{c}{IC~4191} &
\multicolumn{1}{c}{NGC~7027} 
}
\startdata
\multicolumn{5}{l}{Collisionally Excited Lines} \\
N$^o$/H$^+$     & 7.25$^{+0.21}_{-0.16}$ & 6.9$^{....}_{-0.5}$    & 6.4$^{....}_{-0.4}$    & 6.1$^{....}_{-0.3}$    \\ 
N$^+$/H$^+$     & 7.85$^{+0.05}_{-0.06}$ & 7.02$^{+0.16}_{-0.10}$ & 6.74$^{+0.12}_{-0.09}$ & 7.19                   \\
O$^o$/H$^+$     & 7.76$^{+0.06}_{-0.05}$ & 7.68$^{.....}_{-0.14}$ & 7.34$^{.....}_{-0.11}$ & 7.28$^{.....}_{-0.04}$ \\
O$^+$/H$^+$     & 7.24                   & 7.30                   & 7.5$^{+0.3}_{-0.2}$    & 7.68                   \\
O$^{+2}$/H$^+$  & 8.26$^{+0.04}_{-0.05}$ & 8.59$^{+0.04}_{-0.05}$ & 8.79$\pm0.05$          & 8.44                   \\
Ne$^{+2}$/H$^+$ & 7.41$^{+0.03}_{-0.04}$ & 7.58$\pm0.03$         & 7.79$^{+0.03}_{-0.04}$ & \nodata                \\
S$^+$/H$^+$     & 5.5$\pm0.3$            & 5.93                   & 5.67                   & 5.80                   \\
\enddata
\tablenotetext{a}{In units of 12+$\log$(X/H).}
\end{deluxetable} 

%% file: tab5.tex
\begin{deluxetable}{llll}
\rotate
\tablecaption{Atomic Data for Z$>$30 Ions \label{tab5}}
\tabletypesize{\small}
\tablewidth{7.5in}
\tablehead{
\colhead{Ion} &
\colhead{Levels} &
\colhead{Transition Probabilities} &
\colhead{Collision Strengths}
}
\startdata
Br III & ALL 2.05      & \citet{BH86a}   & \nodata \\
       & \citet{BH86a} (2, 3) & & \\
Br IV  & ALL 2.05      & \citet{BH86a}   & \nodata \\
Kr III & ALL 2.05      & \citet{BH86b}   & \citet{Sc97} \\
Kr IV  & ALL 2.05      & \citet{BH86a}   & \citet{Sc97} \\
Kr V   & ALL 2.05      & \citet{BH86a}   & \citet{Sc97} \\
Rb IV  & \citet{PW85}  & \citet{BH86b}   & \nodata \\
Rb V   & \citet{PP84}  & \citet{BH86a}   & \nodata \\
Sr II  & \citet{M52}   & \citet{Br98} (5-1, 5-2)  & \nodata \\
Sr IV  & \citet{HP76}  & \citet{BCH88}   & \nodata \\
Sr V   & \citet{HP74}  & \citet{BH86b}   & \nodata \\
Sr VI  & \citet{PP84}  & \citet{BH86a}   & \nodata \\
Y V    & \citet{RE72}  & \citet{BCH88}   & \nodata \\
Y VI   & \citet{PR86}  & \citet{BH86b}   & \nodata \\
Zr VII & \citet{RA76}  & \citet{BH86b}   & \nodata \\
Te III & \citet{JTD92} & \citet{Bi95}    & \nodata \\
I III  & \citet{TJ93b} & \citet{Bi95}    & \nodata \\
I V    & \citet{KSJ88} & \citet{Bi95}    & \nodata \\
Xe III & \citet{Pe88}  & \citet{Bi95}    & \citet{SB98} \\
Xe IV  & \citet{TJP93} & \citet{Bi95}    & \citet{SB98} \\
Xe V   & \citet{G99}   & \citet{Bi95}    & \nodata \\
Xe VI  & \citet{CJ00}  & \citet{Bi95}    & \citet{SB98} \\
Cs V   & \citet{TJ93a} & \citet{Bi95}    & \nodata \\
Cs VI  & \citet{TJK91} & \citet{Bi95}    & \nodata \\
Ba II  & \citet{KL99}  & \citet{KFW02}   & \citet{SB98} \\
Ba IV  & \citet{Sa93}  & \citet{Bi95}    & \citet{SB98} \\
Ba V   & \citet{Re83}  & \citet{Bi95}    & \nodata \\
Ba VII & \citet{TJ92}  & \citet{Bi95}    & \nodata \\
Ba VIII& \citet{CJG02} & \citet{Bi95}    & \nodata \\
Pb II  & \citet{M58}   & \citet{SSJ05}   & \nodata \\
\enddata
\end{deluxetable}

%% file: tab6.tex
\begin{deluxetable}{ccccclc}
\tablecaption{Emission Line Identifications and Intensities \label{tab6} }
\tablewidth{0pt}
\tablehead{
\colhead{$\lambda^{\rm corr}_{\rm obs}$(\AA)} & \colhead{{\footnotesize FWHM(km\,s$^{-1}$)}}&
\colhead{$F_{\rm obs}$}
& \colhead{$F_{\rm corr}$} &\colhead{S/N} & \colhead{Line Identification} &
\colhead{Notes\tablenotemark{a}}
}
\startdata
&&&{\large \bf IC~2501} \\
3315.23 & 23.73   &  0.0088  &  0.0200  &   7.2& [Mn III] 3315.01& :?             \\
3323.75 & 19.16   &  0.0116  &  0.0260  &  13.9&  Ne II   3323.73&                \\
3327.17 & 27.98   &  0.0087  &  0.0190  &  10.4&  Ne II   3327.15&                \\
3328.69 & 19.13   &  0.0056  &  0.0130  &   8.4&  N  II   3328.72&                \\
3334.83 & 18.57   &  0.0529  &  0.1200  &  50.0&  Ne II   3334.84&                \\
3340.81 & 13.75   &  0.0188  &  0.0420  &  13.3&  N  II   3340.87&                \\
3342.64 & 45.56   &  0.0527  &  0.1200  &  37.0& [Ne III] 3342.54&                \\
3353.22 & 33.40   &  0.0071  &  0.0160  &  11.5& [Cl III] 3353.21&                \\
\nodata & \nodata &  \nodata &  \nodata &  \nodata &  \nodata            & \\
7507.48 & 22.85  &  0.0048  &  0.0027  &  13.6&  N  I    7507.61& :               \\
7509.97 & 57.84  &  0.0033  &  0.0019  &   7.1&                 &\\
7512.69 & 52.14  &  0.0044  &  0.0025  &   7.1&  He I    7513.33&                \\
7514.36 & 36.15  &  0.0038  &  0.0022  &  10.7&  C  III  7514.30&                \\
7515.77 & 22.81  &  0.0173  &  0.0099  &  32.2&  O  III  7715.99& \\
7530.42 & 29.59  &  0.5710  &  0.3300  & 627.6& [Cl IV]  7530.80&                \\
7534.70 & 30.21  &  0.0022  &  0.0013  &   7.1& [Xe IV]  7535.44& \\
\enddata
\tablecomments{The complete version of this table is in the electronic
edition of the Journal.  The printed edition contains only a sample.}
\tablenotetext{a}{: = uncertain identification; ? = uncertain
feature; bl = line blend; ns = blend with night-sky line.}
\end{deluxetable}

%% file: tab7.tex
\begin{deluxetable}%
{l@{\extracolsep{0.1in}}c@{\extracolsep{0pc}}ccc@{\extracolsep{-0.3pc}}c
@{\extracolsep{0.1in}}c@{\extracolsep{0pc}}ccc@{\extracolsep{-0.3pc}}c
@{\extracolsep{0.1in}}c@{\extracolsep{0pc}}ccc@{\extracolsep{-0.3pc}}c
@{\extracolsep{0.1in}}c@{\extracolsep{0pc}}ccc@{\extracolsep{-0.3pc}}c
@{\extracolsep{0.1in}}c@{\extracolsep{0pc}}ccc@{\extracolsep{-0.3pc}}c
@{\extracolsep{0.2pc}}c@{\extracolsep{0pc}}c}
\tablewidth{9.5in}
\setlength{\tabcolsep}{0.014in}
\rotate
\tablecaption{Possible Krypton Line Identifications \label{tab7}}
\tabletypesize{\tiny}
\tablehead{
 & 
\multicolumn{5}{c}{NGC 2440} &
\multicolumn{5}{c}{IC 2501} &
\multicolumn{5}{c}{IC 4191} &
\multicolumn{5}{c}{NGC 7027} &
\multicolumn{5}{c}{IC 418} 
 & & \\
\cline{2-6}\cline{7-11}\cline{12-16}\cline{17-21}\cline{22-26}
 & 
\multicolumn{1}{c}{$\lambda_o$,} &
\multicolumn{1}{c}{$\delta V$\tablenotemark{b}} &
\multicolumn{1}{c}{Mult\tablenotemark{c}} &
\multicolumn{2}{c}{IDI\tablenotemark{d}} &
\multicolumn{1}{c}{$\lambda_o$,} &
\multicolumn{1}{c}{$\delta V$} &
\multicolumn{1}{c}{Mult} &
\multicolumn{2}{c}{IDI} &
\multicolumn{1}{c}{$\lambda_o$,} &
\multicolumn{1}{c}{$\delta V$} &
\multicolumn{1}{c}{Mult} &
\multicolumn{2}{c}{IDI} &
\multicolumn{1}{c}{$\lambda_o$,} &
\multicolumn{1}{c}{$\delta V$} &
\multicolumn{1}{c}{Mult} &
\multicolumn{2}{c}{IDI} &
\multicolumn{1}{c}{$\lambda_o$,} &
\multicolumn{1}{c}{$\delta V$} &
\multicolumn{1}{c}{Mult} &
\multicolumn{2}{c}{IDI} & & \\
\multicolumn{1}{c}{Transition} & 
\multicolumn{1}{c}{I/I$_{H\beta}$\tablenotemark{a}} &
\multicolumn{1}{c}{(km s$^{-1}$)} &
 & & &
\multicolumn{1}{c}{I/I$_{H\beta}$} &
\multicolumn{1}{c}{(km s$^{-1}$)} &
 & & &
\multicolumn{1}{c}{I/I$_{H\beta}$} &
\multicolumn{1}{c}{(km s$^{-1}$)} &
 & & &
\multicolumn{1}{c}{I/I$_{H\beta}$} &
\multicolumn{1}{c}{(km s$^{-1}$)} &
 & & &
\multicolumn{1}{c}{I/I$_{H\beta}$} &
\multicolumn{1}{c}{(km s$^{-1}$)} &
 & & &
\multicolumn{1}{c}{Ref} &
\multicolumn{1}{c}{Notes}
}
\startdata
\textbf{\textrm{[}Kr III\textrm{]} 4p$^4$ $^3$P$_2$--4p$^4$ $^1$D$_2$ \lam6826.70} 
 & 6826.74 &   1.8       & 0/0     & 5A      & &
   6826.82 &   5.3       & 0/0     & 5A      & bl &
   6826.39? & -13.6       & 0/0     & 6B      & : &
   6826.91 &   9.2       & \nodata & \nodata & bl &
   6826.87 &   7.5       & 0/0     & 5A      & bl & & \tablenotemark{e,f} \\ \cline{1-1}
C I 3p $^1$P$_1$--4d $^1$D$^{o}_{2}$ (V21) \lam6828.12 
 & 1.0(-5) & -60.6       & \nodata & \nodata & &
   4.6(-5) & -57.1       & \nodata & \nodata & &
   1.5(-5) & -76.0       & \nodata & \nodata & &
   5.7(-4) & -53.2       & \nodata & \nodata & &
   3.3(-4) & -54.9       & \nodata & \nodata & & B95,Z05 & \\
He I 3s $^3$S$_1$--16p $^3$P$^o$ \lam6827.88                                    
 &         & -50.1       & 0/0     & \nodata & & 
\textit{2.5(-5)}           & -46.6       & 0/0     & \nodata & & 
           & -65.5       & 0/0     & \nodata & &
\textit{4.6(-4)}           & -42.6       & 0/0     & \nodata & bl &
\textit{3.2(-4)}           & -44.4       & \nodata & \nodata & & B95 & \\
\textrm{[}Fe IV\textrm{]} 3d$^5$ $^4$P$_{1/2}$--3d$^5$ $^2$D5$_{3/2}$ \lam6826.50            
 &         &  10.5       & 4/0     & 5A      & &
           &  14.1       & 4/0     & 5A      & &
           &  -4.8       & 4/1     & 3A      & &
           &  18.0       & 4/0     & 7A      & & 
           &  16.3       & 4/0     & 7C      & & & \\
Fe II\textrm{]} c4F$_{7/2}$--y2G$^{o}_{9/2}$ \lam6826.79 
 &         &  -2.2       & 4/0     & 6C      & &
           &  14.1       & 4/0     & 6C      & &
           & -17.6       & 4/0     & 9D      & &
           &   5.3       & 4/0     & 7A      & & 
           &   3.5       & 4/0     & 6B      & & & \\
\enddata 

\tablerefs{B95=PN NGC 7027 \citep{B95}, B00=Orion Nebula
\citep{B00}, E04=Orion Nebula \citep{E04}, H95=PN NGC 6886
\citep{H95}, H01=PN IC 5217 \citep{H01}, P04=PN NGC 5315 \citep{P04},
S03=PN IC 418 \citep{S03}, ZL02=PN Mz 3 \citep{ZL02}, and Z05=PN NGC
7027 \citep{Z05}}

\tablenotetext{a}{Wavelength: 1) $\lambda_o$ are geocentric corrected
wavelength in \AA, 2) ``OUT'' means not in observed range, 3) ``?''
denotes an uncertain feature.  Intensity: 1) numbers in parentheses
are exponents, 2) italics denote corrected intensities attributable
to the $s$-process transition, 3) bracketed
values as upper limits for unobserved features.}
\tablenotetext{b}{Observed ($\lambda_o$)-transition wavelength (km s$^{-1}$).}
\tablenotetext{c}{EMILI multiplet check statistics: number
expected/number observed}
\tablenotetext{d}{EMILI IDI value/rank followed by asterisk (certain ID),
colon (uncertain ID), or ``bl'' (blend).  Definition of IDI given in
Section ~\ref{Z30id}}
\tablenotetext{e}{Unidentified line in IC 418 \citet{B00}.}
\tablenotetext{f}{Unidentified line in Orion Nebula \citet{S03}.} 
\tablenotetext{g}{NGC 2440, IC 4191, NGC 7027: Identified as a separate line.}
\tablenotetext{h}{NGC 7027: Affected strongly by telluric absorption.}
\end{deluxetable}

%% file: tab8.tex
\begin{deluxetable}%
{l@{\extracolsep{0.1in}}c@{\extracolsep{0pc}}ccc@{\extracolsep{-0.3pc}}c
@{\extracolsep{0.1in}}c@{\extracolsep{0pc}}ccc@{\extracolsep{-0.3pc}}c
@{\extracolsep{0.1in}}c@{\extracolsep{0pc}}ccc@{\extracolsep{-0.3pc}}c
@{\extracolsep{0.1in}}c@{\extracolsep{0pc}}ccc@{\extracolsep{-0.3pc}}c
@{\extracolsep{0.1in}}c@{\extracolsep{0pc}}ccc@{\extracolsep{-0.3pc}}c
@{\extracolsep{0.2pc}}c@{\extracolsep{0pc}}c}
\tablewidth{9.5in}
\setlength{\tabcolsep}{0.014in}
\rotate
\tablecaption{Possible Xenon Line Identifications \label{tab8}}
\tabletypesize{\tiny}
\tablehead{
 & 
\multicolumn{5}{c}{NGC 2440} &
\multicolumn{5}{c}{IC 2501} &
\multicolumn{5}{c}{IC 4191} &
\multicolumn{5}{c}{NGC 7027} &
\multicolumn{5}{c}{IC 418} 
 & & \\
\cline{2-6}\cline{7-11}\cline{12-16}\cline{17-21}\cline{22-26}
 & 
\multicolumn{1}{c}{$\lambda_o$,} &
\multicolumn{1}{c}{$\delta V$\tablenotemark{b}} &
\multicolumn{1}{c}{Mult\tablenotemark{c}} &
\multicolumn{2}{c}{IDI\tablenotemark{d}} &
\multicolumn{1}{c}{$\lambda_o$,} &
\multicolumn{1}{c}{$\delta V$} &
\multicolumn{1}{c}{Mult} &
\multicolumn{2}{c}{IDI} &
\multicolumn{1}{c}{$\lambda_o$,} &
\multicolumn{1}{c}{$\delta V$} &
\multicolumn{1}{c}{Mult} &
\multicolumn{2}{c}{IDI} &
\multicolumn{1}{c}{$\lambda_o$,} &
\multicolumn{1}{c}{$\delta V$} &
\multicolumn{1}{c}{Mult} &
\multicolumn{2}{c}{IDI} &
\multicolumn{1}{c}{$\lambda_o$,} &
\multicolumn{1}{c}{$\delta V$} &
\multicolumn{1}{c}{Mult} &
\multicolumn{2}{c}{IDI} & & \\
\multicolumn{1}{c}{Transition} & 
\multicolumn{1}{c}{I/I$_{H\beta}$\tablenotemark{a}} &
\multicolumn{1}{c}{(km s$^{-1}$)} &
 & & &
\multicolumn{1}{c}{I/I$_{H\beta}$} &
\multicolumn{1}{c}{(km s$^{-1}$)} &
 & & &
\multicolumn{1}{c}{I/I$_{H\beta}$} &
\multicolumn{1}{c}{(km s$^{-1}$)} &
 & & &
\multicolumn{1}{c}{I/I$_{H\beta}$} &
\multicolumn{1}{c}{(km s$^{-1}$)} &
 & & &
\multicolumn{1}{c}{I/I$_{H\beta}$} &
\multicolumn{1}{c}{(km s$^{-1}$)} &
 & & &
\multicolumn{1}{c}{Ref} &
\multicolumn{1}{c}{Notes}
}
\startdata
\textbf{\textrm{[}Xe III\textrm{]} 5p$^4$ $^3$P$_2$--5p$^4$ $^1$D$_2$ \lam5846.77} 
 & 5846.73 &   -2.1  & \nodata & \nodata &  & 
   5846.64? &   -6.7  & \nodata & \nodata & : &
   5846.77 &    0.0  & \nodata & \nodata & & 
   5846.65 &   -6.2  & \nodata & \nodata & bl & 
   5846.70 &   -3.6  & 0/0     & 5       & * & & \tablenotemark{e} \\ \cline{1-1}
He II 5--31 \lam5846.66
 & 3.2(-4) &    3.6  & 0/0     & 5B      & * &
   7.7(-6) &   -1.0  & 0/0     & 3A      & & 
   5.2(-5) &    5.6  & 0/0     & 4A      & * &
   4.2(-4) &   -0.5  & 0/0     & 1A      & bl &        
   1.3(-4) &    2.1  & 0/0     & 3A      & & Z05 & \\ % 4-5 8-3,4 11-4 
\textrm{[}Fe II\textrm{]} a2G$_{7/2}$--c2G$_{9/2}$ \lam5847.32 
 &         &  -30.3  & 3/0     & 7C      & & 
           &  -34.9  & 3/0     & 7C      & &
           &  -28.2  & 3/0     & 7C      & &     
   \textit{1.4(-4)}        &  -34.4  & 3/0     & \nodata & & 
           &  -31.8  & 3/0     & \nodata & & & \\ %11-3
Fe II 4d e4F$_{5/2}$--4f 2[3]$^{o}_{5/2}$ \lam5846.78 
 &         &   -2.6  & 0/0     & 3A      & & 
           &   -7.2  & 0/0     & 4B      & &
           &   -0.5  & 0/0     & 4A      & & 
           &   -6.7  & 0/0     & 4B      & & 
           &   -4.1  & 0/0     & 4C      & & & \\
Fe I x5P$^{o}_{1}$--6d 2[3/2]$_1$ \lam5846.60 
 &         &    6.7  & \nodata & \nodata & &
           &    2.1  & \nodata & \nodata & &        
           &    8.7  & \nodata & \nodata & &        
           &    2.6  & \nodata & \nodata & &        
           &    5.1  & 0/0     & 3A      & & & \\
\multicolumn{28}{l}{}\\
\textbf{\textrm{[}Xe IV\textrm{]} 5p$^3$ $^4$S$^{o}_{3/2}$--5p$^3$ $^2$D$^{o}_{5/2}$ \lam5709.20}
 & 5708.95? &   -13.1 & 0/0     & 5       & * &
   5708.97? &   -12.1 & 1/0     & 5       & : &
   5708.98? &   -11.6 & 1/0     & 7       & * & 
   5708.89 &   -16.2 & 1/1      & 7       & * & 
   5708.91? &   -15.2 & 1/1     & 7       & * & & \tablenotemark{e} \\
\enddata

\tablerefs{E04=Orion Nebula \citep{E04}, H01=PN IC 5217 \citep{H01},
P04=PN NGC 5315 \citep{P04}, ZL02=PN Mz 3 \citep{ZL02}, and Z05=PN NGC
7027 \citep{Z05}}

\tablenotetext{a-d}{Same as for Table~\ref{tab7}.}
\tablenotetext{e}{Unidentified line in IC 418 \citep{S03}.}
\tablenotetext{f}{Identified as separate line in all spectra.}
\tablenotetext{g}{On edge of NGC 2440, IC 2501, IC 4191 spectra,
where poorest wavelength calibration is expected.}
\tablenotetext{h}{\protect{L}isted by \citet{H01} as ``unlikely ID''.}
\tablenotetext{i}{IC 4191 and NGC 7027: Identified as separate line.}
\end{deluxetable}

%% file: tab9.tex
\begin{deluxetable}%
{l@{\extracolsep{0.1in}}c@{\extracolsep{0pc}}ccc@{\extracolsep{-0.3pc}}c
@{\extracolsep{0.1in}}c@{\extracolsep{0pc}}ccc@{\extracolsep{-0.3pc}}c
@{\extracolsep{0.1in}}c@{\extracolsep{0pc}}ccc@{\extracolsep{-0.3pc}}c
@{\extracolsep{0.1in}}c@{\extracolsep{0pc}}ccc@{\extracolsep{-0.3pc}}c
@{\extracolsep{0.1in}}c@{\extracolsep{0pc}}ccc@{\extracolsep{-0.3pc}}c
@{\extracolsep{0.2pc}}c@{\extracolsep{0pc}}c}
\tablewidth{9.5in}
\setlength{\tabcolsep}{0.014in}
\rotate
\tablecaption{Possible Bromine Line Identifications \label{tab9}}
\tabletypesize{\tiny}
\tablehead{
 & 
\multicolumn{5}{c}{NGC 2440} &
\multicolumn{5}{c}{IC 2501} &
\multicolumn{5}{c}{IC 4191} &
\multicolumn{5}{c}{NGC 7027} &
\multicolumn{5}{c}{IC 418} 
 & & \\
\cline{2-6}\cline{7-11}\cline{12-16}\cline{17-21}\cline{22-26}
 & 
\multicolumn{1}{c}{$\lambda_o$,} &
\multicolumn{1}{c}{$\delta V$\tablenotemark{b}} &
\multicolumn{1}{c}{Mult\tablenotemark{c}} &
\multicolumn{2}{c}{IDI\tablenotemark{d}} &
\multicolumn{1}{c}{$\lambda_o$,} &
\multicolumn{1}{c}{$\delta V$} &
\multicolumn{1}{c}{Mult} &
\multicolumn{2}{c}{IDI} &
\multicolumn{1}{c}{$\lambda_o$,} &
\multicolumn{1}{c}{$\delta V$} &
\multicolumn{1}{c}{Mult} &
\multicolumn{2}{c}{IDI} &
\multicolumn{1}{c}{$\lambda_o$,} &
\multicolumn{1}{c}{$\delta V$} &
\multicolumn{1}{c}{Mult} &
\multicolumn{2}{c}{IDI} &
\multicolumn{1}{c}{$\lambda_o$,} &
\multicolumn{1}{c}{$\delta V$} &
\multicolumn{1}{c}{Mult} &
\multicolumn{2}{c}{IDI} & & \\
\multicolumn{1}{c}{Transition} & 
\multicolumn{1}{c}{I/I$_{H\beta}$\tablenotemark{a}} &
\multicolumn{1}{c}{(km s$^{-1}$)} &
 & & &
\multicolumn{1}{c}{I/I$_{H\beta}$} &
\multicolumn{1}{c}{(km s$^{-1}$)} &
 & & &
\multicolumn{1}{c}{I/I$_{H\beta}$} &
\multicolumn{1}{c}{(km s$^{-1}$)} &
 & & &
\multicolumn{1}{c}{I/I$_{H\beta}$} &
\multicolumn{1}{c}{(km s$^{-1}$)} &
 & & &
\multicolumn{1}{c}{I/I$_{H\beta}$} &
\multicolumn{1}{c}{(km s$^{-1}$)} &
 & & &
\multicolumn{1}{c}{Ref} &
\multicolumn{1}{c}{Notes}
}
\startdata
\textbf{\textrm{[}Br III\textrm{]} 4p$^3$ $^4$S$^{o}_{3/2}$--4p$^3$ $^2$D$^{o}_{5/2}$ \lam6131.0}
 & \nodata & \nodata & \nodata & \nodata & &
   6130.70 & -14.7   & 1/1     & 4A      & * &
   6130.34 & -32.3   & 1/1     & 6       & * &
   6130.32 & -33.3   & \nodata & \nodata & bl &
   6130.47 & -25.9   & \nodata & \nodata & * & & \tablenotemark{e,f} \\ \cline{1-1}
C III 7h $^{1,3}$H$^o$--16g $^{1,3}$G \lam6130.30
 &[1.7(-5)]& \nodata & \nodata & \nodata & &
   1.7(-5) &  19.6   & 0/0     & 5B      & &
   3.8(-5) &   2.0   & 0/0     & 2A      & &
   1.1(-4) &   1.0   & 0/0     & 2A      & bl &
   2.8(-5) &   8.3   & 0/0     & 2A      & & & \\
\textrm{]}Ni VI\textrm{]} 3d$^5$ $^4$D$_{5/2}$--3d$^5$ $^2$F1$_{7/2}$ \lam6130.40
 &         & \nodata & \nodata & \nodata & &
           &  14.7   & 1/0     & 7       & &
           &  -2.9   & 1/0     & 5D      & &
\textit{7.7(-5)}        &  -3.9   & 1/0     & 4C      & &
           &   3.4   & 1/0     & 4C      & & & \\
\multicolumn{28}{c}{} \\
\textbf{\textrm{[}Br III\textrm{]} 4p$^3$ $^4$S$^{o}_{3/2}$--4p$^3$ $^2$D$^{o}_{3/2}$ \lam6556.4} 
 & 6555.98? & -19.2      & 1/0     & 7       & : &
   6555.92? & -22.0      & 1/1     & 6       & * &
   6555.87? & -24.2      & 1/1     & 6       & * &
   6555.93? & -21.5      & 1/0     & 8       & * &
   \nodata  & \nodata & \nodata & \nodata &   & & \tablenotemark{e} \\ \cline{1-1}
Fe II 4d $^4$P$_{3/2}$--4f $^2$[2]$^{o}_{5/2}$ \lam6555.94  
 & 4.2(-5) &   1.8       & 0/0     & 3       & &
   7.3(-5) &  -0.9       & 0/0     & 2A      & &
   8.5(-5) &  -3.2       & 0/0     & 2A      & &
   4.3(-4) &  -0.5       & 0/0     & 2A      & &
  [2.1(-4)]& \nodata     & \nodata & \nodata & & & \\
\enddata 

\tablerefs{B95=PN NGC 7027 \citep{B95}}

\tablenotetext{a-d}{Same as for Table~\ref{tab7}.}
\tablenotetext{e}{Computed using energy levels from \citet{BH86a}, see text.}
\tablenotetext{f}{Unidentified line in IC 418 \citet{S03}.}
\end{deluxetable}

%% file: tab10.tex
\begin{deluxetable}%
{l@{\extracolsep{0.1in}}c@{\extracolsep{0pc}}ccc@{\extracolsep{-0.3pc}}c
@{\extracolsep{0.1in}}c@{\extracolsep{0pc}}ccc@{\extracolsep{-0.3pc}}c
@{\extracolsep{0.1in}}c@{\extracolsep{0pc}}ccc@{\extracolsep{-0.3pc}}c
@{\extracolsep{0.1in}}c@{\extracolsep{0pc}}ccc@{\extracolsep{-0.3pc}}c
@{\extracolsep{0.1in}}c@{\extracolsep{0pc}}ccc@{\extracolsep{-0.3pc}}c
@{\extracolsep{0.2pc}}c@{\extracolsep{0pc}}c}
\tablewidth{9.5in}
\setlength{\tabcolsep}{0.014in}
\rotate
\tablecaption{Possible Identifications From Other Z$>30$ Ions \label{tab10}}
\tabletypesize{\tiny}
\tablehead{
 & 
\multicolumn{5}{c}{NGC 2440} &
\multicolumn{5}{c}{IC 2501} &
\multicolumn{5}{c}{IC 4191} &
\multicolumn{5}{c}{NGC 7027} &
\multicolumn{5}{c}{IC 418} 
 & & \\
\cline{2-6}\cline{7-11}\cline{12-16}\cline{17-21}\cline{22-26}
 & 
\multicolumn{1}{c}{$\lambda_o$,} &
\multicolumn{1}{c}{$\delta V$\tablenotemark{b}} &
\multicolumn{1}{c}{Mult\tablenotemark{c}} &
\multicolumn{2}{c}{IDI\tablenotemark{d}} &
\multicolumn{1}{c}{$\lambda_o$,} &
\multicolumn{1}{c}{$\delta V$} &
\multicolumn{1}{c}{Mult} &
\multicolumn{2}{c}{IDI} &
\multicolumn{1}{c}{$\lambda_o$,} &
\multicolumn{1}{c}{$\delta V$} &
\multicolumn{1}{c}{Mult} &
\multicolumn{2}{c}{IDI} &
\multicolumn{1}{c}{$\lambda_o$,} &
\multicolumn{1}{c}{$\delta V$} &
\multicolumn{1}{c}{Mult} &
\multicolumn{2}{c}{IDI} &
\multicolumn{1}{c}{$\lambda_o$,} &
\multicolumn{1}{c}{$\delta V$} &
\multicolumn{1}{c}{Mult} &
\multicolumn{2}{c}{IDI} & & \\
\multicolumn{1}{c}{Transition} & 
\multicolumn{1}{c}{I/I$_{H\beta}$\tablenotemark{a}} &
\multicolumn{1}{c}{(km s$^{-1}$)} &
 & & &
\multicolumn{1}{c}{I/I$_{H\beta}$} &
\multicolumn{1}{c}{(km s$^{-1}$)} &
 & & &
\multicolumn{1}{c}{I/I$_{H\beta}$} &
\multicolumn{1}{c}{(km s$^{-1}$)} &
 & & &
\multicolumn{1}{c}{I/I$_{H\beta}$} &
\multicolumn{1}{c}{(km s$^{-1}$)} &
 & & &
\multicolumn{1}{c}{I/I$_{H\beta}$} &
\multicolumn{1}{c}{(km s$^{-1}$)} &
 & & &
\multicolumn{1}{c}{Ref} &
\multicolumn{1}{c}{Notes}
}
\startdata
\textbf{\textrm{[}Rb IV\textrm{]} 4p$^4$ $^3$P$_2$--4p$^4$ $^1$D$_2$ \lam5759.55}
 & 5759.98 &  22.4   & \nodata & \nodata & &
   \nodata & \nodata & \nodata & \nodata & &
   5759.61 &   3.1   & \nodata & \nodata & bl &
   5759.59 &   2.1   & \nodata & \nodata & bl &
   5759.78? &  12.0   & \nodata & \nodata & : & & \tablenotemark{e,f} \\ \cline{1-1}
He II 5--47 \lam5759.74
 & 5.1(-4) &  12.5   & 0/0     & 6B      & * &
  [2.5(-6)]& \nodata & \nodata & \nodata & &
   2.4(-5) &  -6.8   & 0/0     & 4B      & bl &
   3.1(-4) &  -7.8   & 0/0     & 5A      & bl &
   2.0(-5) &   2.1   & 0/0     & 3A      & & Z05 & \\
\textrm{[}Fe II\textrm{]} a2H$_{9/2}$--c2D$_{5/2}$ \lam5759.30
 &         &  35.4   & 0/0     & 5A      & &
           & \nodata & \nodata & \nodata & &
  \textit{1.5(-5)}         &  16.1   & 0/0     & 3A      & &
  \textit{2.0(-4)}         &  15.1   & 0/0     & 6B      & &
           &  25.0   & 0/0     & 5B      & & & \\
\multicolumn{28}{c}{} \\
\textbf{\textrm{[}Rb V\textrm{]} 4p$^3$ $^4$S$^{o}_{3/2}$--4p$^3$ $^2$D$^{o}_{3/2}$ \lam5363.6} 
 & \nodata & \nodata & \nodata & \nodata & &
   5363.81 &  11.7   & \nodata & \nodata & : &
   5363.96? &  20.1   & \nodata & \nodata & : & 
   5363.62 &   1.1   & \nodata & \nodata & * &
   \nodata & \nodata & \nodata & \nodata & & & \\ \cline{1-1}
\textrm{[}Ni IV\textrm{]} 3d$^7$ $^4$F$_{7/2}$--3d$^7$ $^2$G$_{9/2}$ \lam5363.35 
 &[3.4(-5)]& \nodata & \nodata & \nodata & &
   6.7(-6) &  25.7   & 4/0     & 7       & &
   5.4(-6) &  34.1   & 4/0     & 8C      & &
   9.0(-5) &  15.1   & 4/0     & 7B      & &
  [2.3(-5)]& \nodata & \nodata & \nodata & & many & \\
O II 4f.F $^2$[4]$^{o}_{7/2}$--4d$^\prime$ $^2$F$_{7/2}$ \lam5363.80
 &         & \nodata & \nodata & \nodata & &
           &   0.6   & 0/0     & 3A      & &
           &   8.9   & 0/0     & 5A      & &
           & -10.1   & 0/0     & 6A      & &
           & \nodata & \nodata & \nodata & & & \\
\enddata

\tablerefs{Z05=PN NGC7027 \citep{Z05}}

\tablenotetext{a-d}{Same as for Table~\ref{tab7}.}
\tablenotetext{e}{Unidentified line in IC 418 \citet{S03}.}
\tablenotetext{f}{NGC 2440, IC 4191: Contaminated by flare or ghost.}
\end{deluxetable}

%% file: tab11.tex
\begin{deluxetable}{lrrrrrrr@{\extracolsep{0.06in}}r@{\extracolsep{0in}}r}
\tabletypesize{\footnotesize}
\setlength{\tabcolsep}{0.04in}
\tablecaption{Summary of Possible Z$>$30 Ion Line Intensities\tablenotemark{a} \label{tab11}}
\tablehead{
  & \multicolumn{7}{c}{PNe} &  \multicolumn{2}{c}{\ion{H}{2} regions}
  \\ \cline{2-8}\cline{9-10}
  & & & & \multicolumn{3}{c}{NGC~7027} & & & \\ \cline{5-7}
\multicolumn{1}{c}{Transition} &
\multicolumn{1}{c}{NGC~2440} &
\multicolumn{1}{c}{IC~2501} &
\multicolumn{1}{c}{IC~4191} &
\multicolumn{1}{c}{(a)} &
\multicolumn{1}{c}{(b)} &
\multicolumn{1}{c}{(c)} & 
\multicolumn{1}{c}{IC~418} &
\multicolumn{1}{c}{Orion} &
\multicolumn{1}{c}{NGC~3576} 
}
\startdata
\vspace{0.0001in}[Br III] \lam6131.0 & $<$1.7(-5) & \textbf{1.7(-5)} & \textbf{3.8(-5)} & \textbf{7.7(-5)} & \textbf{7.6(-5)} & $<$1.1(-4) & \textbf{2.8(-5)} &    \nodata &    \nodata \\
\vspace{0.0001in}[Br III] \lam6556.4 &    4.2(-5) & \textbf{7.3(-5)} & \textbf{8.5(-5)} & \textbf{4.3(-4)} &    \nodata &    \nodata &   \nodata  &    \nodata &    \nodata \\
\vspace{0.0001in}[Br IV] \lam7368.1  & $<$2.6(-5) & $<$4.4(-5) & $<$3.5(-5) &    4.4(-5) &    2.0(-5) &    \nodata & $<$4.9(-5) &    \nodata &    \nodata \\
\vspace{0.0001in}[Br IV] \lam9450.5  &    \nodata &    \nodata &    \nodata &    7.5(-5) &   $<$7(-6) &    \nodata & $<$5.3(-5) &    \nodata &    \nodata \\
\vspace{0.0001in}[Kr III] \lam6826.70 &    \nodata & \textbf{2.5(-5)} & 1.5(-5) & \textbf{4.6(-4)}  & \textbf{4.1(-4)} &    4.4(-4) & \textbf{3.2(-4)} &    7.0(-5) & $<$3.0(-5) \\
\vspace{0.0001in}[Kr IV] \lam5346.02  & \textbf{3.5(-4)} & \textbf{6.1(-5)} & \textbf{2.4(-4)} & \textbf{1.9(-3)} & \textbf{1.9(-3)} &    1.5(-3) & \textbf{3.5(-5)} &    \nodata &    \nodata \\
\vspace{0.0001in}[Kr IV] \lam5867.74  & \textbf{4.6(-4)} & \textbf{8.5(-5)} & \textbf{1.9(-4)} & \textbf{2.6(-3)} & \textbf{2.6(-3)} &    2.3(-3) & \textbf{3.5(-5)} &    7.8(-5) & $<$5.0(-5) \\
\vspace{0.0001in}[Kr IV] \lam6107.8  &    5.6(-6) & $<$4.3(-6) &    \nodata & \textbf{7.2(-5)} & \textbf{6.3(-5)} &      5(-5) & $<$1.4(-5) &    \nodata &    \nodata \\
\enddata
\tablerefs{NGC~7027=(a) present measurement; (b) \citet{PB94}; (c) \citet{Z05}; Orion=\citet{B00}; NGC~3576=\citet{GR04}}
\tablenotetext{a}{In units of I(H$\beta$)=1, certain identifications in bold.}
\end{deluxetable}

%% file: tab12.tex
\begin{deluxetable}{llllll@{\extracolsep{0.06in}}l@{\extracolsep{0in}}l}
\setlength{\tabcolsep}{0.03in}
\tabletypesize{\footnotesize}
\tablecaption{$s$-process Ionic/Elemental Abundances\tablenotemark{a} \label{tab12}}
\tablewidth{0pt}
\tablehead{
  & \multicolumn{5}{c}{PNe} &  \multicolumn{2}{c}{\ion{H}{2} regions} \\
\cline{2-6}   \cline{7-8}
\colhead{\textit{X}$^{+i}$/H$^+$/Transition}
& \colhead{NGC~2440}
& \colhead{IC~2501} & \colhead{IC~4191}
& \colhead{NGC~7027} & \colhead{IC~418}
& \colhead{Orion} & \colhead{NGC~3576} }
\startdata
Br$^{+2}$/H$^+$ \lam6131.0 & $<$1.77                & 2.21$^{+0.16}_{-0.19}$ & 2.18$^{+0.14}_{-0.17}$ & 2.52$^{+0.09}_{-0.10}$ & 2.41$^{+0.17}_{-0.22}$ & \nodata & \nodata \\  
Br$^{+2}$/H$^+$ \lam6556.4 & 2.13$^{+0.19}_{-0.28}$ & 2.70$^{+0.18}_{-0.22}$ & 2.44$^{+0.16}_{-0.21}$ & 3.14$^{+0.10}_{-0.12}$ & $<$3.07                & \nodata & \nodata \\
Adopted                  & 2.13$^{+0.19}_{-0.28}$ & 2.27$^{+0.14}_{-0.15}$ & 2.24$^{+0.11}_{-0.13}$ & 2.57$^{+0.08}_{-0.09}$ & 2.41$^{+0.17}_{-0.22}$ & \nodata & \nodata \\
Br$^{+3}$/H$^+$ \lam7368.1 & $<$2.21                & $<$2.89                & $<$2.74                & 2.53$^{+0.09}_{-0.11}$ & $<$3.00                & \nodata & \nodata \\
Br$^{+3}$/H$^+$ \lam9450.5 & \nodata                & \nodata                & \nodata                & 2.74$^{+0.11}_{-0.14}$ & $<$3.02                & \nodata & \nodata \\
.............. Adopted   & $<$2.21                & $<$2.89                & $<$2.74                & 2.58$^{+0.08}_{-0.09}$ & $<$3.00                & \nodata & \nodata \\
Kr$^{+2}$/H$^+$ \lam6826.70 & \nodata                & 2.4$^{+0.3}_{-0.5}$    & 1.89$^{+0.13}_{-0.17}$ & 3.39$^{+0.07}_{-0.08}$ & 3.53$^{+0.13}_{-0.16}$ & 3.01    & $<$2.47 \\
Kr$^{+3}$/H$^+$ \lam5346.02 & 2.96$^{+0.09}_{-0.10}$ & 2.75$^{+0.12}_{-0.15}$ & 3.28$\pm0.10$          & 3.78$^{+0.02}_{-0.03}$ & 2.60$^{+0.17}_{-0.18}$ & \nodata & \nodata \\
\enddata
\tablenotetext{a}{In units of 12+$\log$(N/H).}
\tablenotetext{b}{Ar abundances using $t^2=0.00$.}
\end{deluxetable}